\pdfoutput=1
\documentclass[conf]{new-aiaa} % for conference papers
\usepackage[utf8]{inputenc}
\usepackage{textcomp}

\usepackage{graphicx}
\usepackage{amsmath}
\usepackage[version=4]{mhchem}
\usepackage{siunitx}
\usepackage{longtable,tabularx}
\usepackage{multirow, makecell}

\usepackage{multicol}

\usepackage{tikz}
\usetikzlibrary{shapes.geometric, arrows}

\usepackage[utf8]{inputenc}
\usepackage[T1]{fontenc}

\usepackage{graphicx}
\usepackage{subcaption}
	\graphicspath{{./figs/}}
\usepackage{float}
\usepackage{epstopdf}

\usepackage{wrapfig}% embedding figures/tables in text (i.e., Galileo style)
\usepackage{threeparttable}% tables with footnotes
\usepackage{dcolumn}% decimal-aligned tabular math columns
	\newcolumntype{d}{D{.}{.}{-1}}
\usepackage{nomencl}% automatic nomenclature generation via makeindex
	\makeglossary

\usepackage{fancyvrb}% extended verbatim environments
	\fvset{fontsize=\footnotesize,xleftmargin=2em}

\usepackage{ifthen}
\usepackage{xstring}

\usepackage{lettrine}% dropped capital at beginning of paragraph
\hypersetup{
	colorlinks=true,
	citecolor=blue
}

\newcommand{\expandref}[2][]{%
	\IfStrEq{#2}{e}{Equation#1}{%
	\IfStrEq{#2}{f}{Figure#1}{%
	\IfStrEq{#2}{c}{Chapter#1}{%
	\IfStrEq{#2}{s}{Section#1}{%
	\IfStrEq{#2}{t}{Table#1}{%
	\IfStrEq{#2}{fs}{Fig#1.}{%
	\IfStrEq{#2}{ss}{Sec.}{%
	\IfStrEq{#2}{es}{Eq#1.}{%
	\IfStrEq{#2}{ap}{Appendix#1}{%
	}}}}}}}}}
}

\newcommand{\eref}[1]{\hyperref[#1]{\expandref{e}(\ref*{#1})}} % equation ref
\newcommand{\fref}[1]{\hyperref[#1]{\expandref{f}\ref*{#1}}} % figure ref
\newcommand{\tref}[1]{\hyperref[#1]{\expandref{t}\ref*{#1}}} % table ref
\newcommand{\cref}[1]{\hyperref[#1]{\expandref{c}\ref*{#1}}} % chapter ref
\newcommand{\sref}[1]{\hyperref[#1]{\expandref{s}\ref*{#1}}} % section ref
\newcommand{\sfref}[1]{\hyperref[#1]{(\protect\subref*{#1})}} % subfig ref
\newcommand{\figref}[1]{\hyperref[#1]{\expandref{fs}\ref*{#1}}} % fig. ref
\newcommand{\secref}[1]{\hyperref[#1]{\expandref{ss}\ref*{#1}}} % sec. ref
\newcommand{\eqnref}[1]{\hyperref[#1]{\expandref{es}\ref*{#1}}} % eq. ref
\newcommand{\appref}[1]{\hyperref[#1]{\expandref{ap}\ref*{#1}}} % eq. ref

\newcommand{\multiref}[3]{%
\IfStrEq{#1}{e}{\expandref[s]{#1}(\ref{#2}) -- (\ref{#3})}{%
\expandref[s]{#1}\ref{#2} -- \ref{#3}
}}

\newcommand{\degreei}{\ensuremath{^\circ}} % Internal command for the degree symbol
\newcommand{\degree}{\degreei{}} % Degree symbol. Use this
\newcommand{\degrees}{\ensuremath{^\circ ~}} % Degree symbol with a forced nbsp
\newcommand{\makecase}[2]{ %
\IfStrEq{#1}{u}{\texorpdfstring{\MakeUppercase #2}{#2}}{%
\IfStrEq{#1}{l}{\texorpdfstring{\MakeLowercase #2}{#2}}{%
}}
}

\newcommand{\figwidth}{0.8\textwidth}

\usepackage{ar} % The aspect ratio package for the AR symbol.
\usepackage{siunitx} % A standard way to include SI units in text

\newcommand{\ar}{\AR} % Just an alias to the aspect ratio command because I can never remember if it's \ar or \AR. Now it's both

\newcommand{\aeffi}{\ensuremath{\alpha_\text{eff}}}
\newcommand{\aeff}{\aeffi{}}

\usepackage{mathtools}

\def\etal.{et\penalty50\ al.}

\newcommand{\diff}[2]{\frac{\text{d}#1}{\text{d}#2}}

\usepackage{diagbox}

\newcommand{\mnote}[1]{}
\newcommand{\revnote}[2]{#1}

\title{A Low-Order Method for Prediction of Separation and Stall on Unswept Wings}

\author{Pranav Hosangadi \footnote{PhD. Candidate, Dept. of Mechanical and Aerospace Engineering, Campus Box 7910. Student Member, AIAA.}
and Ashok Gopalarathnam\footnote{Professor, Dept. of Mechanical and Aerospace Engineering, Campus Box 7910. Associate Fellow, AIAA}}
\affil{North Carolina State University, Raleigh, NC 27695-7910}
\makenomenclature

\begin{document}

\maketitle

\newcommand{\methodname}{low-order method~}
\newcommand{\methodabbr}{LOM~}

\newcommand{\aefftarget}{\ensuremath{\alpha_\text{eff}^\text{target}}}
\newcommand{\aeffprime}{\ensuremath{\alpha_\text{eff}'}}

\newcommand*{\doi}[1]{doi: \url{#1}}

\begin{abstract}
A low-order method is presented for aerodynamic prediction of wings operating at near-stall and post-stall flight conditions. The method is intended for use in design, modeling, and simulation. In this method, the flow separation due to stall is modeled in a vortex-lattice framework as an effective reduction in the camber, or ``decambering.'' For each section of the wing, a parabolic decambering flap, hinged at the separation location of the section, is calculated through iteration to ensure that the lift and moment coefficients of the section match with the values from the two-dimensional viscous input curves for the effective angle of attack of the section. As an improvement from earlier low-order methods, this method also predicts the separation pattern on the wing. Results from the method, presented for unswept wings having various airfoils, aspect ratios, taper ratios, and small, quasi-steady roll rates, are shown to agree well with experimental results in the literature, and computational solutions obtained as part of the current work.

\end{abstract}

%010-introduction.tex
\nomenclature{$\Gamma$}{Circulation strength of vortex}

% 020-background.tex
\nomenclature{$\hat{n}$}{Unit normal vector}
\nomenclature{$[AIC]$}{Aerodynamic influence coefficient matrix}
\nomenclature{$(t/c)_\text{max}$}{Maximum airfoil thickness as a fraction of chord}

%030-methodology.tex
\nomenclature{$\alpha$}{Angle of attack}
\nomenclature{$\aeff$}{Effective angle of attack of wing section}
\nomenclature{$\alpha_{0L}$}{Zero-lift angle of attack of airfoil}
\nomenclature{$\delta_l$}{Inclination of the decambering flap at hinge point}
\nomenclature{$f$}{Separation point location as a fraction of chord}
\nomenclature{$m$}{Height of the decambering flap at the trailing edge as a fraction of chord}
\nomenclature{$x$}{Chordwise coordinate, positive towards trailing edge}
\nomenclature{$y$}{Spanwise coordinate, positive towards right wingip}
\nomenclature{$z$}{Vertical coordinate, positive towards upper surface}
\nomenclature{$C_l, C_L$}{Lift coefficient of airfoil, wing}
\nomenclature{$C_d, C_D$}{Drag coefficient of airfoil, wing}
\nomenclature{$C_m, C_M$}{Coefficient of pitching moment about quarter chord (airfoil), root-quarter-chord (wing)}
\nomenclature{$C_n$}{Normal force coefficient of airfoil}
\nomenclature{$\omega$}{Angular velocity}
\nomenclature{$V_\infty$}{Freestream velocity}

%040-results.tex
\nomenclature{$\ar$}{Aspect ratio, $b^2/S$}
\nomenclature{$b$}{Wingspan}
\nomenclature{$c$}{Chord}
\nomenclature{$S$}{Wing planform area}

%041-geomtable.tex
\nomenclature{$\lambda$}{Taper ratio, root chord / tip chord}
\nomenclature{$pb/2V$}{Nondimensionalized roll rate}
\nomenclature{$Re$}{Reynolds number based on $c_\text{mean}$}

%042-naca4415-other.tex
\nomenclature{$c_\text{mean}$}{\revnote{Mean geometric chord}{\#3.6}}

% \input 010-introduction.tex
% \input 020-background.tex
% \input 030-methodology.tex
% \input 040-results.tex
% \input 100-conclusions.tex

% \section*{Nomenclature}
% \begin{multicols}{2}
    % \renewcommand{\nomname}{}
    % {\setstretch{1} \printnomenclature}
% \end{multicols}

{\printnomenclature}

% \printnomenclature

% \lipsum[1-5]

\section{Introduction}

Aircraft normally operate in the ``linear region'' of aerodynamics.
This region, which occurs at low angles of attack, is characterized by mostly attached flow, and a linear variation of lift with angle of attack.
In the linear region, the boundary layer is thin and the flow can be approximated by a potential-flow solution.
The behavior of airfoils and wings at low angles of attack has been thoroughly studied, and extensive data on the forces and moments acting on lifting surfaces in the linear region is available from a variety of experimental, numerical, and theoretical sources \cite{abbott,stuttgarter_profilkatalog_1,lsats_3,xfoil,drela_avl}.%\cite{abbott,stuttgarter_profilkatalog_1,lsats_1,lsats_2,lsats_3,lsats_4,xfoil}.
As the angle of attack increases, an adverse pressure gradient forms on the upper surface of the airfoil/wing. The adverse pressure gradient %
% reduces the velocity of the flow and
causes the boundary layer to thicken and then separate from the surface.
The thick, separated boundary layer changes the effective shape of the body. The flow can no longer be approximated by the potential-flow theory, and
the lift produced drops in comparison to the linear curve.
As the angle of attack increases further, the adverse pressure gradient intensifies, and the location at which the flow separates moves forward towards the leading edge.
Beyond a limiting angle of attack ($\alpha_\text{stall}$), the lift produced starts to decrease with an increasing angle of attack. The drop in lift is accompanied by a significant increase in drag and a drop in pitching moment, and the airfoil/wing is said to have stalled.

Although a majority of applications operate in the linear region, post-stall aerodynamics are commonly experienced by applications such as wind turbines, helicopters, and even some fixed-wing aircraft. A solid understanding of near-stall and post-stall flows is crucial to the success of these applications.
Aerodynamic models that can be used to rapidly predict the loads acting on wings and aircraft configurations have applications in preliminary design, flight dynamics characterization, and flight simulation.
Due to the requirement for rapid predictions, low-order models are especially useful in such applications.
Low-order predictive methods based on potential flow, such as the vortex lattice method (VLM), are well established in predicting the force and moment characteristics, and spanwise distributions of the forces and moments, on wings and multiple-surface configurations at low angles of attack, where the flow can be approximated by potential flow.
The development of the first steady VLM dates back to work done by Hedman in the 1960s
\cite{Hedman_VLM_1965}, with unsteady modifications introduced by Thrasher et al. \cite{Thrasher1977} and Konstadinopoulos et al. \cite{Konstadinopoulos1985} in the 1970s and 80s.
However, the VLM, with various modifications
and enhancements, is used even today for low-order modeling and
engineering applications, with recent examples ranging from flight
dynamics analysis \cite{drela_avl, Obradvoic_Subbarao_2011}, analysis of
yacht sails \cite{FIDDES199635}, calculation of aerodynamic
interference effects \cite{Elzebda_Mook_Nayfeh_1994, Rossow_1995,
Karkehabadi_2004}, post-stall analysis \cite{Mukherjee_poststall_2006,
Rom1993}, flapping-wing analysis
\cite{Nguyen_insect_UVLM_2016, Fritz_Long_2004, Stanford_Beran_2010,Hirato2019},
wind turbines \cite{SIMOES1992129, PESMAJOGLOU20001}, design
optimization \cite{CUSHER201435, Stanford_Beran_2010, Mariens2014} and
aeroelasticity \cite{MURUA201246, Palacios_Murua_Cook_2010,
Murua_Palacios_Graham_2012}.
Modified VLMs have also been extensively used for modeling steady and unsteady flows past delta-wings \cite{Traub1999},
propeller aerodynamics \cite{Kobayakawa1985},
propeller-wing interactions \cite{Witkowski1989},
ground effect and formation flight \cite{Frazier2003, King2005,Han2005,Zhang2017},
compressibility effects and transonic flow over wings \cite{Batina1986,Melin2010},
system identification \cite{Venkataraman2019},
and for rapid performance prediction in adaptive control of aircraft \cite{Kim2010,Menon2013}.
The current work, along similar lines, aims to extend the VLM for modeling separation and stall.

Extensive research has been carried out to extend the range of potential-flow-based methods to obtain aerodynamic predictions beyond the linear region. Some methods \cite{Valarezo1994,Phillips2007} use empirical relations based on the lift curve of the airfoil obtained from experimental or CFD data to obtain maximum wing lift. While these methods can accurately predict $C_{L,\text{max}}$, they do not predict the wing behavior well in the post-stall region.
Another common approach is
to modify the potential flow-based equations of traditional low-order methods to
model the effects of thick and separated boundary layers. {Often, this
modification is achieved using a strip-theory based approach. Strip theory has been widely used to predict the behavior of wings based on the behavior of their airfoils \cite{Rodden1959,PamadiTaylor-1984-JofAC-SpinningAirplane,Liu1988,Cebeci1989,Wang2010,Castellani2017}.} To calculate the loads on the wing, it
is discretized into strips and the behavior of each strip is approximated to \revnote{that}{\#3.16} of the corresponding airfoil.
For each airfoil, viscous input data is supplied, often in the form of airfoil
lift ($C_l$-$\alpha$) curves which form the convergence criteria while solving the
3-D potential flow equations to calculate spanwise loading. Convergence is achieved by iteratively modifying the circulation distribution over the surface
\cite{tani_1934,schairer_1939,sivells_neely_1947,sears_1956,levinsky_1976,piszkin_levinsky_1976,anderson_llt_1980,mccormick_nonlinear_llt_1989}
($\Gamma$-correction methods),
or the effective angle of attack of the strips \cite{Purser1951,Hunton1953,tseng_lan_1988,bruce_owens_nonlinear_weissinger,van_dam_nonlin_wing_2001,wickenheiser_garcia_2011} ($\alpha$-correction methods).
These approaches yield sufficiently accurate results for simple unswept
geometries, providing a significant cost-benefit compared to higher fidelity
approaches such as CFD.
Dias \cite{Dias2016} uses the Kirchhoff-Helmholtz formulation to obtain the coefficient of lift for each section of a wing represented as a lifting line. The equation of the lifting line is modified to include the effect of separation. The location of the separation point, denoted in that work by $X$, is the variable used to change the viscous behavior of each section. The variation of the location of the separation point with angle of attack is specified as an empirical equation derived by fitting experimental observations. Iterations are performed until the change in the effective angle of attack of the sections due to a change in $X$ becomes negligible.
Chreim et al. \cite{Chreim2018} model viscous effects in their implementation of lifting-line theory by moving the location of the collocation points points in the chordwise direction for each section. Changing the location of the collocation points has the effect of changing the lift-curve slope for each section. The method calculates the required lift-curve slope for each section so that its operating point may fall on the viscous lift curve of the airfoil.
Gabor et al. \cite{Gabor2016} apply a $\Gamma$-correction to a vortex lattice method to calculate the circulation distribution required on the surface to change the strip behavior to be identical to that of an airfoil. Corrections to the circulation strength of each vortex ring are obtained using a Jacobian-based Newton iteration.
The work by dos Santos and Marques \cite{Santos2018} uses a $\Gamma$-correction approach to apply viscous corrections to inviscid solutions obtained from a VLM. The elements of the aerodynamic influence coefficient (AIC) matrix are corrected based on Kirchhoff's model for separated flow over a flat plate, where the separation point location is estimated using a semi-empirical model developed by Leishman and Beddoes \cite{Leishman1989}.
Kharlamov et al. \cite{Kharlamov2018} use a 2D URANS solver modified to obtain solutions for ``infinite-swept-wings'', which includes the effects of sweep in the lift-curves of the 2D sections of the wing. An $\alpha$-correction method is used to modify the effective angle of attack of the sections of the wing. The correction to $\alpha$ for each section is based on the change in $C_l$ required at that section and the lift-curve slope.
A similar approach is used by Gallay and Laurendeau \cite{Gallay2016} and Parenteau et al. \cite{Parenteau2018,Parenteau2018b}.

A method developed at NCSU's Applied Aerodynamics Group uses the concept of ``decambering'', wherein the camber of the sections of the wing is reduced at high angles of attack to model the separation of the boundary layer and the accompanying reduction in lift.
As with the other methods described above, %readily available
viscous lift data for the airfoil from experiments or computations is supplied to the decambering method. In contrast to the methods discussed previously, a ``decambering flap'' is used to implement the viscous correction by modifying the shape of the effective body.
The decambering approach provides accurate predictions at high angles of attack \cite{Mukherjee_poststall_2006,Paul_Gopa_Iteration_Schemes,gopalarathnam_paul_petrilli_ASM_2012}.
% Motivated by the success of the decambering approach in predicting the lift behavior for unswept wings at high angles of attack, the current research effort aims to extend this success to predicting the post-stall behavior of the lift, drag, and moment on wings having arbitrary planforms, and on multiple-surface configurations.

This paper describes the concept of decambering and its application to a potential-flow method to obtain viscous load predictions for airfoils and wings experiencing separated flow. A novel decambering approach dubbed ``nonlinear decambering'' is presented. In contrast to previous ``linear'' decambering approaches which used two linear decambering flap deflections hinged at predetermined locations to obtain the required drop in lift and moment associated with boundary-layer separation, the nonlinear decambering approach achieves this using a single parabolic decambering flap. The nonlinear flap for each section is hinged at the predicted location of flow separation, allowing the flap to better approximate the shape of the separated boundary layer.
\revnote{The use of a vortex lattice method allows for calculation of the chordwise distribution of the surface loading over the entire lifting surface as opposed only the spanwise distribution that is obtained using an approach based on lifting-line theory. Additionally, the vortex lattice method can correctly calculate the inviscid circulation distribution over swept wings and more complicated planforms, a capability that lifting-line theory does not possess. This capability allows for prediction of post-stall aerodynamics for swept wings, an early version of which is presented in \cite{Hosangadi2015}.}{\#3.2}
The benefits of the nonlinear decambering over other post-stall low-order methods discussed above, including the linear decambering approaches, are the capability to predict separation patterns along the wing span and cross-sectional separated-flow profiles. These benefits serve as essential stepping stones to extension of the current work to predictions of swept-wing stall and viscous wakes behind stalled wings. \revnote{Although the decambering method is incorporated in a VLM in the current work, it can also be applied to other inviscid prediction methods like surface panel methods.}{\#3.7}

An overview of the underlying vortex lattice method (VLM) implemented in this work is given in \sref{sec:vlm}. \sref{sec:decambering} discusses the background of the decambering method and the main assumptions of the approach. A detailed description of the nonlinear decambering approach is given in \sref{sec:nonlin-decambering}. For flow over an airfoil, the application of nonlinear decambering is relatively straightforward, as described in \sref{sec:nld-2D}.
% \multiref{s}{sec:nld-3D}{sec:iter-nld}
\sref{sec:nld-3D} covers the complications involved in applying nonlinear decambering to a three-dimensional wing. Finally, results from the low-order method for various geometries are compared against experimental results % from Ostowari and Naik \cite{naik_ostowari_nrel}
and against 3D RANS CFD solutions obtained using ANSYS Fluent in \sref{sec:unswept-results}.

 \section{The Vortex Lattice Method}
\label{sec:vlm}
The vortex lattice method (VLM) is a numerical method used to solve the three-dimensional potential-flow lifting-surface problem. Its primary advantage over simpler methods such as the Weissinger method or lifting-line theory is that it represents the actual camber shape of the wing, and therefore can be used for a wide variety of cambered wings and planforms. The implementation used in this work is described in detail by Katz and Plotkin \cite{katz_plotkin_book_1991}.

\begin{figure}[!htbp]
    \centering
    \includegraphics{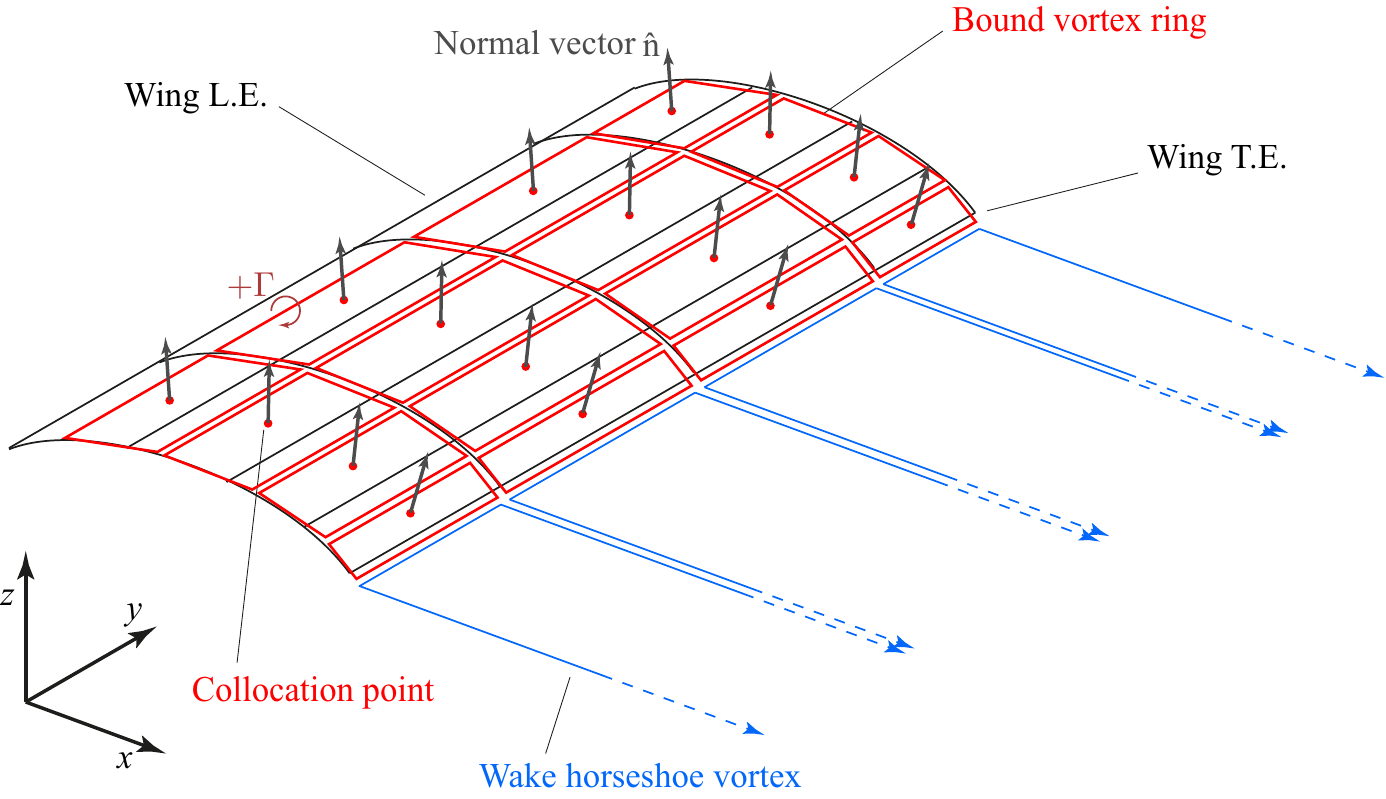}
    \caption{The lattice of vortex ring elements used to discretize a lifting surface in the VLM}
    \label{fig:vlm-panels}
\end{figure}

The geometry is first condensed into a camber surface, which is then discretized into a lattice of panels in the chordwise and spanwise directions. A vortex ring element is assigned to each panel, such that the leading segment of the vortex ring lies at the quarter-chord point of the panel. The trailing segment coincides with the leading segment of the next vortex ring. A collocation point is defined at the three-quarter chord line of the panel. The normal vector to the camber surface at the collocation point ($\hat{n}$) is used to enforce the boundary condition of zero normal flow through the surface. In the current implementation, a steady wake model is used -- the wake is assumed to be flat (no roll-up) and fixed (no change due to angle of attack), and extends to downstream infinity. This wake shape is modeled using horseshoe vortices to discretize the wake. An illustration of the discretization is shown in \fref{fig:vlm-panels}.

At each collocation point, the %Dirichlet
boundary condition of zero normal flow through the camber surface is imposed. This boundary condition can be expressed as:
\begin{align}
    \left(\sum \vec{V}\right) \cdot \hat{n} = 0 \label{eqn:boundary-condition}
\end{align}

\noindent where $\sum \vec{V}$ is the vector sum of all velocities acting at the collocation point. This includes the incoming wind ($\vec{V}_\infty$), the velocities induced at the collocation point due to the vorticity bound to the wing and shed in the wake, velocities induced by any free vortices or vorticity bound to a different surface, and any velocity due to rotation of the body itself.

Since the geometry remains unchanged even with varying angles of attack, the calculation can be simplified using an aerodynamic influence coefficient matrix, $[AIC]$. With the geometry discretized into $M$ chordwise and $N$ spanwise panels, the AIC matrix becomes a square matrix of order ($M \times N$), with each element $a_{i,j}$ specifying the influence on the collocation point of the $i$th panel of the $j$th bound vortex ring.

\begin{align}
    [AIC] \left\{ \Gamma \right\} &= \left\{ RHS \right\} \label{eqn:vortex-lattice-eqn}
\end{align}

The RHS is a known column vector, with each element $r_{i}$ denoting the velocity due to all velocities not arising due to vortex-lattice influences, including $V_\infty$, rotational velocities. Solving \eref{eqn:vortex-lattice-eqn} yields a vector containing the circulation strengths of each vortex ring, which can be used to calculate the loads on the wing using the Kutta-Joukowski theorem.
The lift calculated using the VLM is corrected for thickness effects using an empirical equation given by Katz and Plotkin \cite{katz_plotkin_book_1991}.
\begin{align}
    C_{l, \text{corrected}} = \left[1 + 0.77 (t/c)_\text{max}\right] C_{l}
\end{align}

\section{The Decambering Method}
\label{sec:decambering}
% Potential-flow methods accurately predict the lift and moment characteristics of an airfoil at low angles of attack. As the angle of attack increases, viscous effects start to dominate and potential-flow methods no longer predict the loads correctly. At high angles of attack, the formation of an adverse pressure gradient on the upper surface causes the boundary layer to thicken and then separate from the surface.
The %resulting
change in the shape of the effective body due to boundary-layer separation at high angles of attack causes a reduction in the camber of the airfoil (``decambering''), leading to a drop in lift and moment associated with stall.
The decambering method \cite{Mukherjee_poststall_2006} developed in previous research at NCSU models the reduction in camber at high angles of attack using a linear ``decambering flap'' at the trailing edge of the airfoil, hinged at a fixed chordwise location, along with another decambering flap hinged at the leading edge.
\revnote{On applying the decambering flap, the zero-lift $\alpha$ of the airfoil is changed, with the lift-curve slope remaining unchanged.}{\#3.9}
A potential flow method can then be used to calculate the loads on the modified airfoil. 

The decambering method is easily applied to three-dimensional wings using a strip-theory approach. The wing is divided into chordwise strips with each strip assumed to behave like an airfoil. Decambering flaps are applied to each strip so as to fulfil the following conditions:

\label{sec:decambering-conditions}
\emph{Condition 1:} There is no normal flow through the strip, achieved by imposing the zero-flow boundary condition  normal to the decambered geometry at the collocation points of the vortex lattice. This condition is enforced by solving the linear system of the VLM to obtain the correct circulation strengths for the bound and wake vortices.

\emph{Condition 2:} The operating points for each strip after decambering, given by ($\aeff, C_l$) and ($\aeff$, $C_m$), fall on the $C_l$-$\alpha$ and $C_m$-$\alpha$ curves of the airfoil. This condition is satisfied by the deflection of the decambering flap.

An iterative process is used to enforce both conditions simultaneously. This approach has been shown to satisfactorily predict the lift generated by finite wings \cite{Mukherjee_poststall_2006,Paul_Gopa_Iteration_Schemes,gopalarathnam_paul_petrilli_ASM_2012}. The following section gives a detailed illustration of the modified nonlinear decambering procedure presented in the current work.

 \section{Nonlinear Decambering}
\label{sec:nonlin-decambering}
\begin{figure}[!ht]
    \centering
    \includegraphics[width=3in]{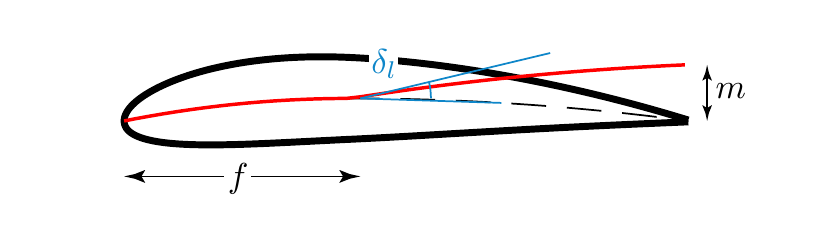}
    \caption{The parabolic flap used by the nonlinear decambering method, hinged at $x/c = f$, with an inclination of $\delta_l$ at the hinge location and a height $m$ at the trailing edge}
    \label{fig:nld-flap-illustration}
\end{figure}

The nonlinear decambering method \cite{Narsipur2018} was developed as an improvement to the linear decambering method, and uses a single parabolic flap as shown in \fref{fig:nld-flap-illustration} to model the reduction in camber due to flow separation at high angles of attack. The nonlinear decambering flap is hinged at the separation location ($f$), and has an inclination ($\delta_l$) at the hinge, and a height ($m$) at the trailing edge. The value of $f$ varies from 0 (separation occurs at the leading edge, i.e. flow is fully separated) to 1 (flow is fully attached, separation occurs at the trailing edge).
In contrast to previously published decambering approaches \cite{Mukherjee_poststall_2006,Paul_Gopa_Iteration_Schemes,gopalarathnam_paul_petrilli_ASM_2012}, this method requires a single flap to account for deviation in both lift and moment from their respective inviscid values.

Non-dimensionalizing the $x$ and $z$ coordinates with chord, we write $\bar{x} = x/c$ and $\bar{z} = z/c$.
The shape of the parabolic decambering flap is given by:
\begin{align}
    \bar{z}\left(\bar{x}\right) = \begin{cases}
    0 & \bar{x} < f \\
    A\bar{x}^2 + B\bar{x} + D & \bar{x} \geq f
    \end{cases} \label{eqn:decambering-camberline}
\end{align}

\noindent where,
\begin{align}
    A &= \frac{m - (1 - f)\tan(\delta_l)}{(1 - f)^2} \label{eqn:decambering-A} \\
    B &= \tan(\delta_l) - 2 A f \label{eqn:decambering-B}  \\
    D &= m - (A + B) \label{eqn:decambering-D}
\end{align}

Using thin-airfoil theory, the change in lift and moment caused by a nonlinear decambering flap having the decambering parameters ($f, \delta_l, m$) are given by:
\begin{align}
    \left[ \begin{array}{c}
         \Delta C_l \\
         \Delta C_m
    \end{array}\right] &= \left[ \begin{array}{cc}
        a_1 & b_1 \\
        a_2 & b_2
    \end{array} \right] \left[ \begin{array}{c}
         A  \\
         B
    \end{array} \right] \label{eqn:decambering-dclcm}
\end{align}

\noindent where,
\begin{align*}
    a_1 &= 3 \theta_f - 3 \pi - 4 \sin \theta_f + \frac{1}{2} \sin 2\theta_f \\
    a_2 &= \frac{3}{4} \sin \theta_f - \frac{3}{8} \sin 2\theta_f + \frac{1}{12} \sin 3\theta_f - \frac{\theta_f}{4} + \frac{\pi}{4} \\
    b_1 &= 2\theta_f - 2\pi - 2 \sin \theta_f \\
    b_2 &= \frac{1}{2} \sin \theta_f - \frac{1}{4} \sin 2\theta_f \\
    \theta_f &= \arccos(1 - 2f)
\end{align*}

% \section{Input Data for the Nonlinear Decambering Method}
The decambering parameters ($f, \delta_l, m$) for each flow condition are calculated using viscous lift, moment, and separation location data obtained from steady two-dimensional CFD solutions or experiments. In case the separation location is not available in the input datasets, Beddoes' modification \cite{Beddoes1983} to the Kirchhoff-Helmholtz solution for separated flow over a flat plate can be used to estimate the location of the separation point as follows:
\begin{align}
    f &= \left(2\sqrt{\frac{C_l}{2\pi \sin(\alpha - \alpha_{0L})}} - 1\right)^2 \label{eqn:beddoes-f}
\end{align}

\section{Nonlinear Decambering Applied to Two-Dimensional Airfoils}
\label{sec:nld-2D}

\begin{figure*}[ht!]
    \centering
    \begin{subfigure}[t]{0.45\textwidth}
        \centering
        \includegraphics[width=2.5in]{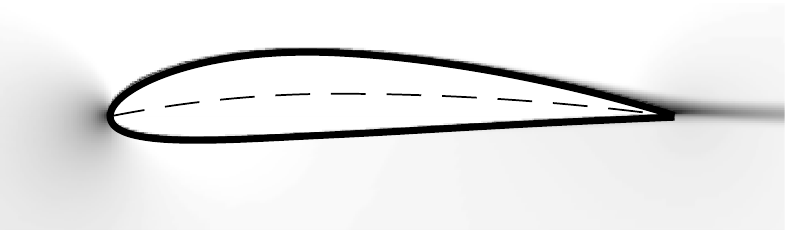}
        \caption{$\alpha = 2\degree$, no decambering\label{sfig:nld-lowalpha}}
    \end{subfigure}
    ~
    \begin{subfigure}[t]{0.45\textwidth}
        \centering
        \includegraphics[width=2.5in]{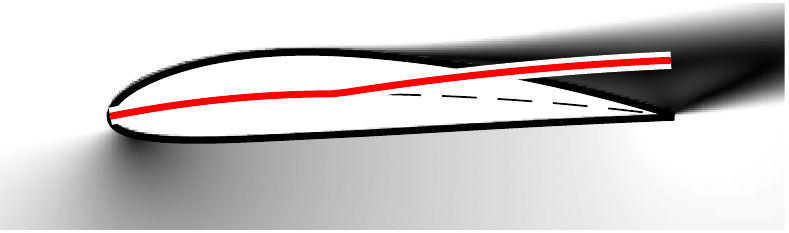}
        \caption{$\alpha = 18\degree$, Nonlinear decambering\label{sfig:nld-highalpha}}
    \end{subfigure}%
    ~
    \begin{subfigure}[t]{0.45\textwidth}
        \centering
        \includegraphics[width=2.5in]{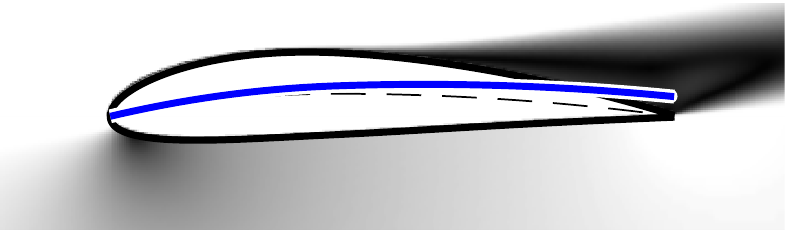}
        \mnote{\#2.2}
        \caption{$\alpha = 18\degree$, linear decambering\label{sfig:nld-linhighalpha}}
    \end{subfigure}
    \caption{The original camberline (black, dashed) and the camberline after nonlinear decambering (red) and linear decambering from Ref. \cite{Paul_Gopa_Iteration_Schemes} (blue) for the NACA 4415 airfoil, overlaid on the CFD-predicted contour of $|\vec{V}|/|\vec{V}_\infty|$\label{fig:nld-illustration}}
\end{figure*}

\begin{figure}[h!t]
    \centering
    \includegraphics[width=\figwidth]{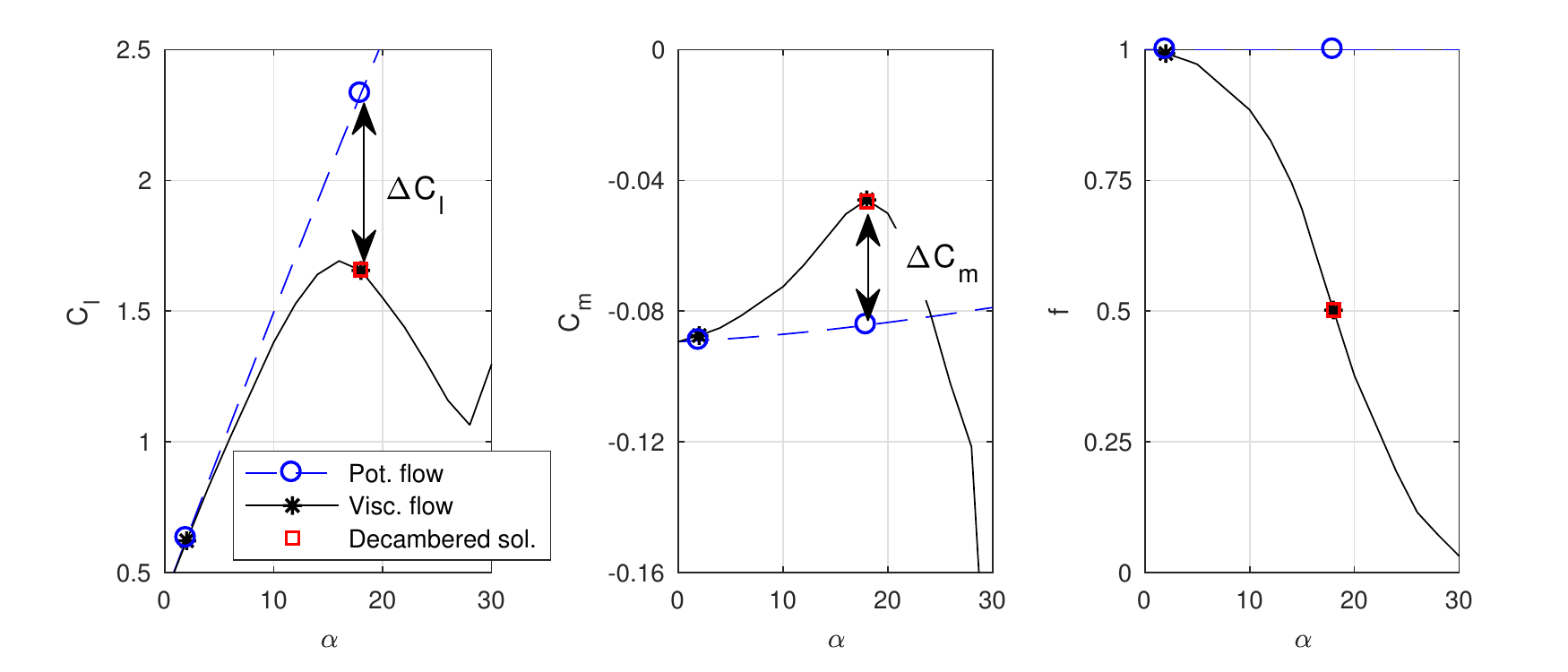}
    \caption{Potential-flow and viscous-flow curves (lines) and operating points (symbols) for the NACA 4415 airfoil}
    \label{fig:nld-targets}
\end{figure}

The concept of nonlinear decambering is easily illustrated using the example of 2D flow past an airfoil shown in  \multiref{f}{fig:nld-illustration}{fig:nld-targets}. At low angles of attack, the boundary layer is fully attached to the surface of the airfoil, as seen in \fref{sfig:nld-lowalpha}.
The resulting loads on the airfoil, shown in \fref{fig:nld-targets} are accurately predicted by an inviscid method and no decambering is required.
As the angle of attack increases, the separated boundary layer causes a deviation in lift and moment from the potential-flow predictions.
From the viscous input data supplied, we obtain the location of the separation point ($f$), and the ``target'' $C_l$ and $C_m$ values that the airfoil is known to produce in viscous flow, marked in \fref{fig:nld-targets} by black asterisks. The required change in $C_l$ and $C_m$ are written as $\Delta C_l = C_{l,\text{viscous}} - C_{l,\text{potential}}$ and $\Delta C_m = C_{m,\text{viscous}} - C_{m,\text{potential}}$, respectively. The separation location and required $\Delta C_l$ and $\Delta C_m$ thus found from the viscous input data are plugged in to \multiref{e}{eqn:decambering-A}{eqn:decambering-dclcm} to obtain the values of the decambering parameters. At low angles of attack where the viscous input curves indicate mostly attached flow, the required $\Delta C_l$ and $\Delta C_m$ are small, and $f \approx 1$. In such cases, using the correct value of $f$ leads to large, non-physical decambering flap deflections to achieve even the small deviations in lift and moment. This problem is avoided by restricting the value of $f$ to a maximum of $0.8$. %This is necessary because a higher value of $f$ would require a large, non-physical decambering flap deflection to obtain the even small $\Delta C_l$ and $\Delta C_m$ that are required in such cases.
The camberline of the airfoil is modified according to \eref{eqn:decambering-camberline}, and the potential-flow solution for the modified camberline is seen to fall on the viscous operating curve of the airfoil.
\revnote{From \fref{sfig:nld-highalpha}--\ref{sfig:nld-linhighalpha}, it can be seen that the nonlinear decambering flap mimics the shape of the separated boundary layer more closely than the linear decambering flap described in Ref. \cite{Paul_Gopa_Iteration_Schemes}.}{\#2.2}

\section{Nonlinear Decambering Applied to Three-Dimensional Wings}
\label{sec:nld-3D}
Similar to the other variations of decambering, nonlinear decambering can be applied to three-dimensional wings using a strip-theory approach. The wing is divided into chordwise strips, with a set of decambering parameters assigned to each strip. These parameters are determined individually for each strip based on the change in lift and moment required at that strip to simultaneously satisfy both conditions mentioned in \secref{sec:decambering} above.

\subsection{Effective Angle of Attack}
\label{sec:aeff}

The angle of attack ($\alpha$) is defined as the angle between the chordline of the airfoil/wing and the incoming wind.
However, the trailing vortex sheet behind the wing induces a downwash at every section of the wing that causes a reduction in the angle of attack experienced by the section. Since the section ``sees'' the incoming wind impinging on itself at this angle, the behavior of the section is governed by this ``effective'' angle of attack ($\aeff$) rather than the wing angle of attack ($\alpha$). Calculation of this effective angle of attack correctly at each section of the wing is crucial to accurate prediction of the aerodynamic loads on the sections, and therefore those on the wing.
%This is achieved by finding the angle of attack at which a two-dimensional section having the same shape and decambering parameters as the wing-section experiences the same normal force as the wing-section.
The effective angle of attack, however, is not an output of a VLM solution.
To calculate the effective angle of attack for a wing section from the VLM solution for a given angle of attack, the same decambering flap applied to the section is also applied to a two-dimensional airfoil. Next, a Newton-Raphson iteration is used to vary the angle of attack of the airfoil until it produces the same normal force ($C_n$) as the wing section. The angle of attack for the 2D airfoil is the effective angle of attack of the wing section.

\subsection{Profile Drag}
Although the VLM can predict induced drag on three-dimensional wings excellently, it is ill-suited to predicting the profile drag of the wing because it is based on potential flow theory. However, the strip theory approach can be used to estimate the profile drag without needing to calculate it by solving the equations of viscous flow. Since the $C_d$ for the airfoil is typically available in viscous airfoil datasets, each strip is assigned a viscous $C_d/C_l$ vs. $\alpha$ curve obtained from the $C_d$ and $C_l$ of the airfoil. When the \methodname solves for the lift distribution on the sections, the $C_d$ for each section is obtained by interpolating the $C_d/C_l$ vs. $\alpha$ curve at the $\aeff$ of the section. The forces and moments resulting from the interpolated $C_d$ for each strip are then added to the total values calculated by the VLM.

\subsection{Decambering Trajectory}
\begin{figure}
    \centering
    \includegraphics[width=4.5in]{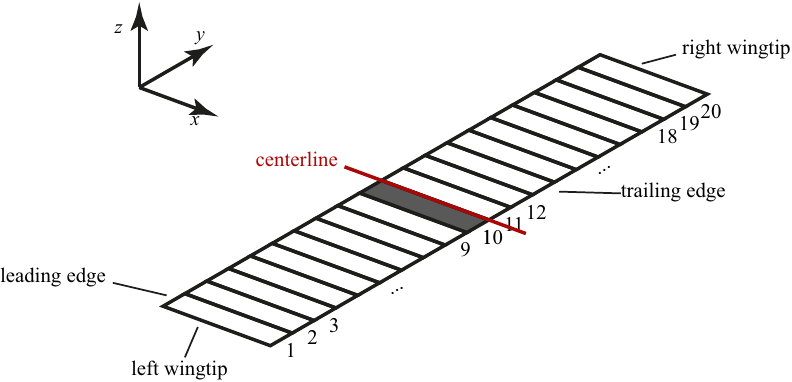}
    \caption{Illustration of the wing geometry divided into strips}
    \label{fig:strips-illustration-3D}
\end{figure}

\begin{figure*}[ht!]
    \centering
    \begin{subfigure}[t]{0.40\textwidth}
        \centering
        \includegraphics[width=\textwidth]{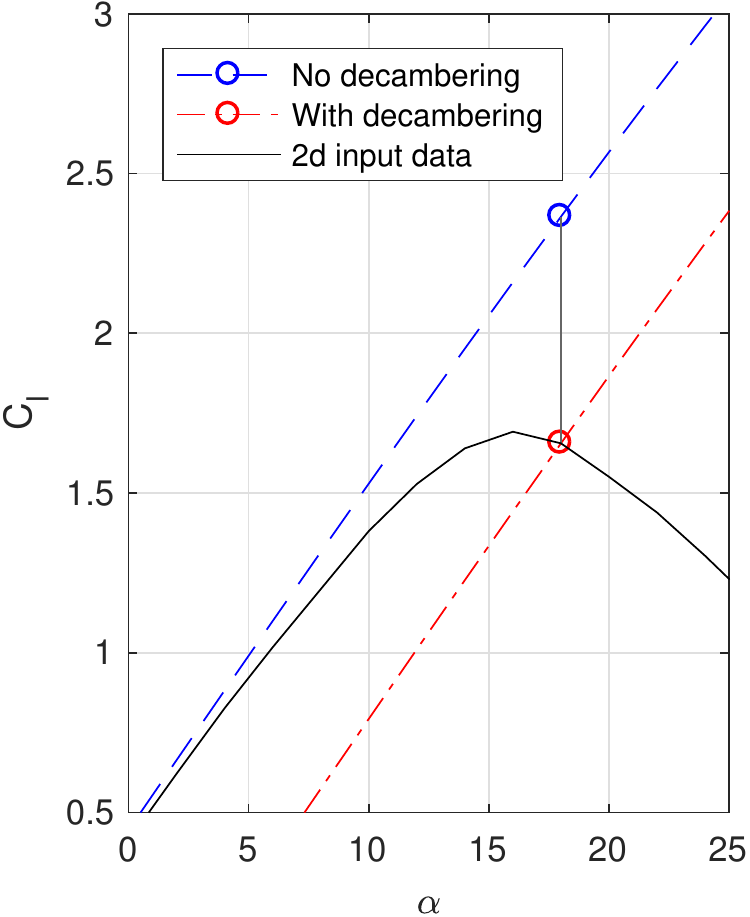}
        \caption{2D airfoil\label{sfig:traj-illustration-2D}}
    \end{subfigure}%
    ~
    \begin{subfigure}[t]{0.40\textwidth}
        \centering
        \includegraphics[width=\textwidth]{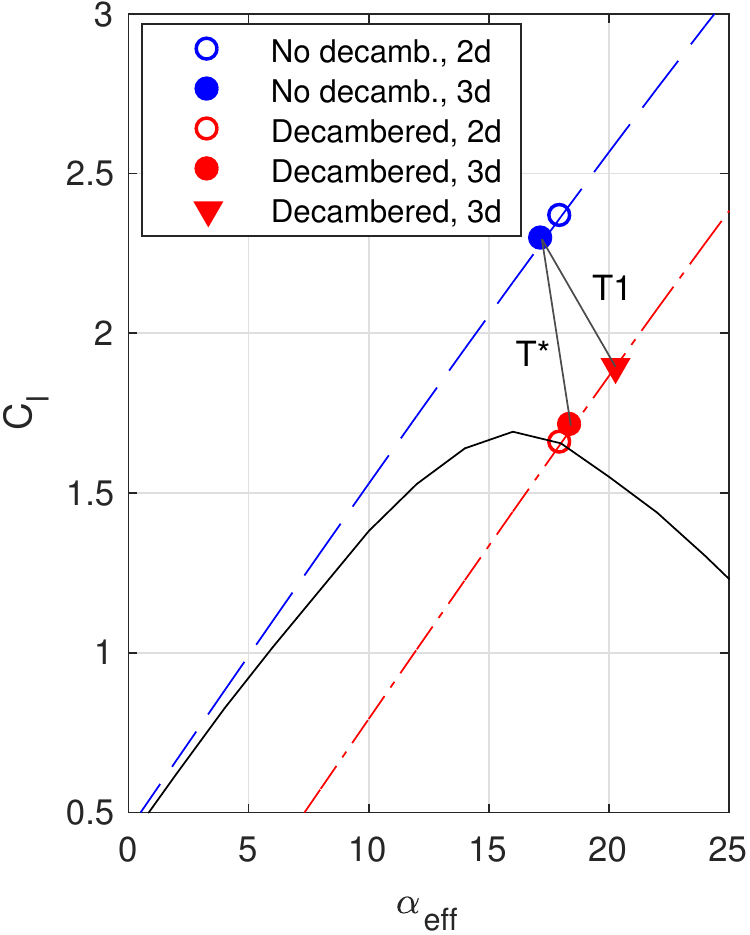}
        \caption{Section of a 3D wing\label{sfig:traj-illustration-3D}}
    \end{subfigure}
    \caption{An illustration of the decambering trajectories for an airfoil and a section of a 3D wing\label{fig:traj-illustration}}
\end{figure*}

In earlier works \cite{Mukherjee_poststall_2006,Paul_Gopa_Iteration_Schemes}, it was shown that the induced flow at a section is affected by the decambering at every other section.
Using the example wing geometry in \fref{fig:strips-illustration-3D} in which the wing span is discretized into 20 strips, the effective angle of attack at some section (section 10, for example) will depend on the downwash at that section. Because this downwash depends on the lift distribution over the span, the effective angle of attack at section 10 depends on the decambering at all the sections. %As a result, the effective angle of attack of a section depends on the decambering parameters of all other sections in addition to the downwash induced by the wingtip vortices.
This behavior is not seen in the case of a 2-dimensional airfoil.
To illustrate this phenomenon, consider the \revnote{effect of a single iteration of decambering on a}{\#3.11} NACA4415 airfoil at an angle of attack of $18\degree$.
Without decambering, the operating point $(\alpha, C_l)$ falls on the inviscid lift curve ($C_l = 2\pi \sin(\alpha - \alpha_{0L})$) for the airfoil, denoted by the dashed blue line in \fref{sfig:traj-illustration-2D}.
\revnote{In the first iteration of decambering, the values of $f$, $\delta_l$, and $m$ are calculated}{\#3.11} using \multiref{e}{eqn:decambering-A}{eqn:decambering-dclcm} above. The airfoil produces less lift ($C_{l,d}$) than in the case without decambering.
The operating point must fall on the lift curve of the decambered airfoil, denoted by the red line in \fref{fig:traj-illustration}.
Now, the effective angle of attack of the airfoil is found by locating the $\alpha$-coordinate at which the lift curve of the decambered airfoil gives a coefficient of lift equal to $C_{l,d}$.
The operating point of the airfoil is observed to have moved vertically downwards as a consequence of decambering (\fref{sfig:traj-illustration-2D}).

Now consider the behavior of a single section (section 10) near the root of the three-dimensional wing shown in \fref{fig:strips-illustration-3D}, \revnote{again undergoing a single decambering iteration}{\#3.11}.
Due to induced downwash effects, the section produces less lift than the airfoil even without any decambering applied, which moves the inviscid (no-decambering) operating point to the filled blue circle in \fref{sfig:traj-illustration-3D}. Now, let us apply the calculated decambering parameters to this section only, and no decambering at any other sections of the wing.
\revnote{The loss in lift on that strip due to the decambering results in strong trailing vortices at the edges of the strip (due to sudden increase in lift going from this strip to the adjacent one). These trailing vortices cause upwash which partly negates the effect of the decambering. The result is that the loss in $C_l$ is not as much as that seen with just the 2D decambering. Calculating the new lift and $\aeff$ of this section as described above, we see that the operating point moves along the trajectory labeled ``T1'' in \fref{sfig:traj-illustration-3D} to the point denoted by the filled red triangle.
Instead, if the requisite decambering is applied to all sections of the wing instead of a single section, there are no strong trailing vortices at the edges of the strip. The upwash due to the entire-wing decambering is much smaller than with the single-strip decambering.
Because of this, the loss in $C_l$ is only slightly smaller than that from the 2D decambering. The operating point now follows a different trajectory labeled ``T*'' to the new operating point denoted by the filled red circle.}{\#3.10}

It is seen from this exercise that the movement of the operating point of a section in the $C_l$ vs. $\aeff$ space depends on the induced flow at the section, which in turn depends on the decambering at every other section of the wing.
The slopes of the trajectory lines for each section are required in order to obtain accurate target viscous operating points from the viscous input curves.
To obtain these slopes, the inviscid solution is calculated to obtain the starting operating point ($\aeff^0, C_{l}^0$) for each section.
The trajectory for each section is assumed to be vertical, as seen for the airfoil, and the intersection of the vertical trajectory with the viscous input curve gives an initial $C_{l}^\text{target}$ and $C_{m}^\text{target}$ for the decambering method. The decambering required at each section is calculated, and the perturbed operating points ($\aeff^p, C_{l}^p$) are found. Now, the trajectory slope at each section is given by:
\begin{align}
    \diff{C_l}{\aeff} &= \frac{C_l^p - C_l^0}{\aeff^p - \aeff^0}
\end{align}

% \begin{figure}[!h]
%     \centering
%     \includegraphics[width=2.75in]{figs/eps_fig/traj-moreiters.eps}
%     \caption{Decambering trajectories for all sections of a 3-dimensional wing remain fairly constant}
%     \label{fig:traj-moreiters}
% \end{figure}

\begin{figure*}[!h]
    \begin{subfigure}[t]{0.45\textwidth}
        \centering
        \includegraphics[width=2.75in]{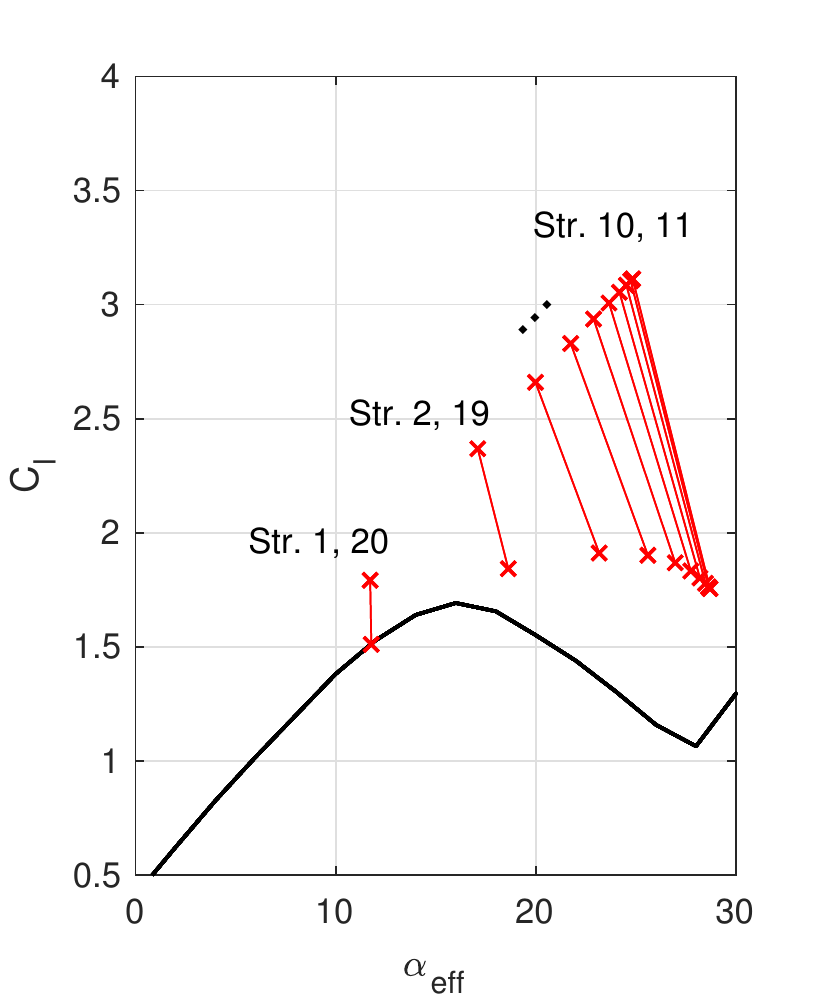}
        \caption{}
        \label{fig:n4415-ar12-swp0-traj30}
    \end{subfigure}%
    ~
    \begin{subfigure}[t]{0.45\textwidth}
        \centering
        \includegraphics[width=2.75in]{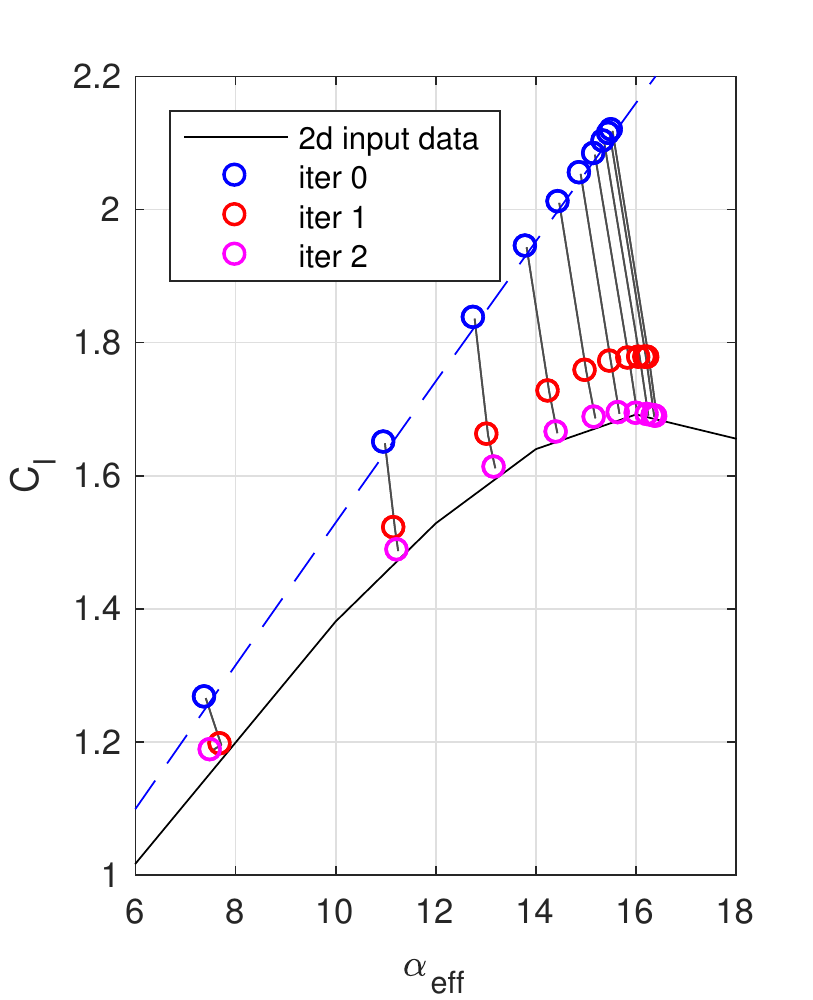}
        \caption{}
        \label{fig:traj-moreiters}
    \end{subfigure}%
    ~
    \caption{(a) Decambering trajectories calculated at a post-stall angle of attack $\alpha = 30\degree$, (b) Decambering trajectories do not change significantly as the iteration advances}
\end{figure*}

\revnote{It was observed that the actual trajectories of the operating points due to decambering were more or less identical to the decambering trajectories calculated in the first iteration at a high angle of attack where all sections require some decambering.
For all results presented in this paper, the trajectories were calculated at $\alpha = 30\degree$.
These trajectories are shown in \fref{fig:n4415-ar12-swp0-traj30} for the NACA4415 $\ar{12}$ wing.
Since the wing is symmetric about the $y$-plane, trajectories for sections $i$ and $20-i+1$ coincide.
During subsequent iterations of the decambering method, the
operating points appear to move in a straight line in the $C_l$-$\aeff$ space due to the trajectories remaining invariant.
The similarity of decambering trajectories between subsequent iterations is shown in \fref{fig:traj-moreiters}. This observation is in agreement with the results of Paul and Gopalarathnam~\cite{Paul_Gopa_Iteration_Schemes}, and allows the method to select an accurate target viscous operating point immediately after the inviscid solution is calculated.}{\#3.11}

\subsection{Iterative Calculation of Decambering Parameters}
\label{sec:iter-nld}
Once the decambering trajectories have been calculated at a preset angle of attack, the \methodname can be used to calculate the wing loads at any $\alpha$. The iterative procedure used to converge on a solution satisfying the conditions from \secref{sec:decambering-conditions} is described below with the aid of the flowchart in \fref {fig:flowchart}.
We start with the decambering parameters for all sections initialized to $f=1, \delta_l = 0, m = 0$, which corresponds to the inviscid solution for flow over the wing.

\tikzstyle{startstop} = [rectangle, rounded corners, minimum width=3cm, minimum height=1cm,text centered, draw=black, fill=white] %
\tikzstyle{process} = [rectangle, minimum width=3cm, minimum height=1.5cm, text centered, draw=black, fill=white]%
\tikzstyle{decision} = [diamond, minimum width=3cm, minimum height=1cm, text centered, draw=black, fill=white, yshift=-0.5cm] %
\tikzstyle{arrow} = [thick,->,>=stealth] %
%
% \begin{figure}[ht]
%     \centering
%     \begin{tikzpicture}[node distance=2cm]
%         \node (start) [startstop] {Start};
%         \node (vlmsolve) [process, below of=start] {Solve VLM};
%         \node (aeff) [process, below of=vlmsolve] {Calculate $\aeff$};
%         \node (targets) [process, below of=aeff] {Obtain viscous targets};
%         \node (converged) [decision, below of=targets, text width=1.5cm] {Solution \\ converged?};
%         \node (modgeom) [process, right of=targets, xshift=2cm] {Modify geometry};
%         \node (calcdec) [process, right of=converged, xshift=2cm, text width=2cm] {Calculate decambering ($f, \delta_l, m$)};
%         \node (stop) [startstop, below of=converged, yshift=-0.5cm] {Stop};
%     \end{tikzpicture}
%     \caption{Flowchart of the \methodname}
%     \label{fig:flowchart}
% \end{figure}
%
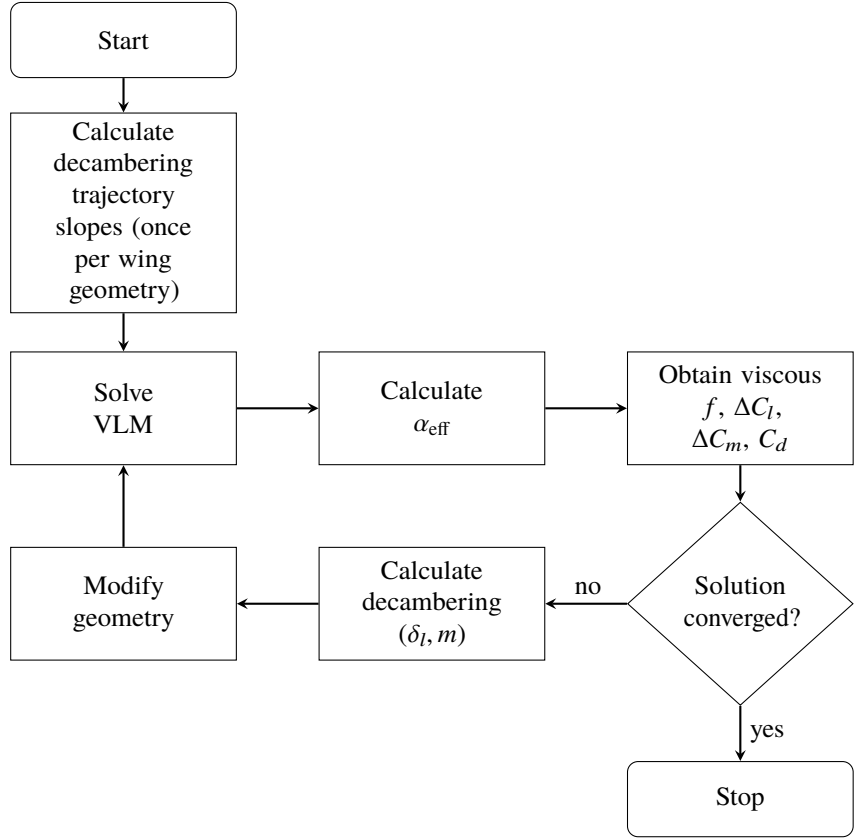
\begin{figure}[ht]
    \centering
    \begin{tikzpicture}[node distance=2.1cm, text width=1.5cm]
        \node (start) [startstop] {Start};
        \node (traj) [process, below of=start, yshift=-0.2cm, text width=2cm] {Calculate decambering trajectory slopes (once per wing geometry)};
        \node (vlmsolve) [process, below of=traj, yshift=-0.5cm] {Solve VLM};
        \node (aeff) [process, right of=vlmsolve, xshift=2cm] {Calculate $\aeff$};
        \node (targets) [process, right of=aeff, xshift=2cm, text width=2.2cm] {Obtain viscous $f$, $\Delta C_l$, $\Delta C_m$, $C_d$};
        \node (converged) [decision, below of=targets, text width=1.5cm] {Solution \\ converged?};
        \node (calcdec) [process, left of=converged, xshift=-2cm, text width=2cm] {Calculate decambering ($\delta_l, m$)};
        \node (modgeom) [process, left of=calcdec, xshift=-2cm] {Modify geometry};
        \node (stop) [startstop, below of=converged, yshift=-0.5cm] {Stop};
        \draw [arrow] (start) -- (traj);
        \draw [arrow] (traj) -- (vlmsolve);
        \draw [arrow] (vlmsolve) -- (aeff);
        \draw [arrow] (aeff) -- (targets);
        \draw [arrow] (targets) -- (converged);
        \draw[arrow] (converged) -- node[anchor=south,xshift=0.6cm] {no} (calcdec);
        \draw[arrow] (converged) -- node[anchor=west] {yes} (stop);
        \draw[arrow] (calcdec) -- (modgeom);
        \draw[arrow] (modgeom) -- (vlmsolve);
    \end{tikzpicture}
    \caption{Flowchart of the \methodname}
    \label{fig:flowchart}
\end{figure}%

\subsubsection{Potential-Flow Solution}\label{step:vlm-solution}
The RHS of the linear system of the VLM is calculated using the normal vectors of the geometry, the velocity of the incoming wind ($V_\infty$), and the rotational velocity (if any) of the wing. Control surface deflections are accounted for by tilting the appropriate normal vectors by the required angle. For the $i$th panel, the new normal vector is given by
\begin{align}
    \hat{n}_{i}^{'} &= \hat{n}_i + \Delta \hat{n}_{i,\text{decambering}} + \Delta\hat{n}_{i,\text{control}}
\end{align}
\noindent where $\hat{n}_i$ is the original normal vector of the panel, with $\Delta \hat{n}_{i,\text{decambering}}$ and  $\Delta\hat{n}_{i,\text{control}}$ denoting the change in normal vector due to decambering and control surface deflection respectively.
The total velocity $\vec{V}_i$ at the collocation point for this panel depends upon $\vec{V}_\infty$ and the velocity produced at the collocation point $\vec{p}_i$ due to rotation about the center of rotation $\vec{p}_\text{rot}$.
\begin{align}
    \vec{V}_i &= \vec{V}_\infty - \vec{\omega} \times (\vec{p}_i - \vec{p}_\text{rot})
\end{align}
Finally, the normal component of the velocity at the collocation point is obtained from the dot product of the total velocity $\vec{V}_i$ with the effective normal vector $\hat{n}_{i}^{'}$.
\begin{align}
    RHS_i &= -\vec{V}_i \cdot \hat{n}_{i}^{'}
\end{align}

The linear system of the vortex lattice is then solved to obtain the potential flow solution.

\subsubsection{Calculation of Effective Angle of Attack}\label{step:aeff-calc}
For each section of the wing, a 2D discrete vortex solver is initialized with a geometry identical to the wing-section, including any alterations due to control surface deflection or decambering. Additional velocities due to the rotation of the wing are not included, since the effect of these velocities will be included in the calculated $\aeff$.
%\PHNote{Need to think of a better way to explain this}.
A Newton iteration is set up to calculate the $\alpha$ required for the 2D solver to produce the same normal force ($C_n$) as the section of the 3D wing. This calculated $\alpha$ is the $\aeff$ for the section.

\subsubsection{Obtaining Target Viscous Coefficients Using Decambering Trajectories}\label{step:traj-target}
\begin{figure}[ht]
    \centering
    \includegraphics[width=2.75in]{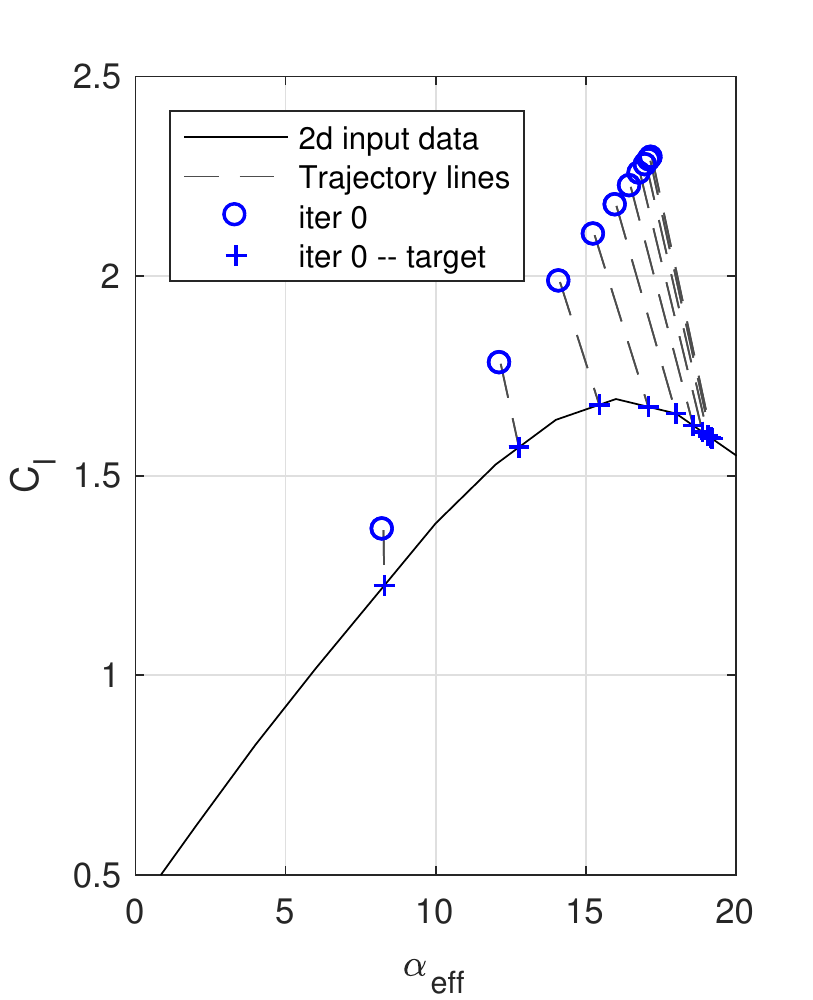}
    \caption{Decambering targets for each section}
    \label{fig:iter0-targets}
\end{figure}

At each section, the target viscous $C_l$ is obtained by finding the intersection of its decambering trajectory line with the pre-calculated slope and the viscous input $C_l$ vs. $\alpha$ curve. \fref{fig:iter0-targets} shows the operating points and decambering targets for the zeroth iteration for each section of the NACA4415 \ar{12} wing at $\alpha=18\degree$.

\subsubsection{Checking for Convergence}\label{step:convergence}
At this stage, the error in $C_l$ and $C_m$ for each section is calculated.

\begin{align}
    \Delta C_{l,i} &= C_{l,i}^\text{target} - C_{l,i} \\
    \Delta C_{m,i} &= C_{m,i}^\text{target} - C_{m,i}
\end{align}

Convergence is said to have been achieved when the mean absolute error in both $C_l$ and $C_m$ are below a specified tolerance, which indicates that the conditions specified in \secref{sec:decambering-conditions} have been fulfilled.
% The results presented in \secref{sec:unswept-results} were obtained using a tolerance of 0.05 for $C_l$ and 0.01 for $C_m$.
Tolerances of 0.05 for $C_l$ and 0.01 for $C_m$ were found to yield sufficiently accurate results, as shown in \secref{sec:unswept-results} below.

\subsubsection{Calculation of Decambering Parameters}\label{step:calc-decam}
If the required tolerance is not met, the increment in decambering needed to achieve the required change in lift and moment is calculated using \multiref{e}{eqn:decambering-A}{eqn:decambering-dclcm}. If a decambering flap already exists at a section prior to the current iteration, the calculated $\delta_l$ and $m$ are added to the preexisting decambering parameters.

\subsubsection{Solution Update}\label{step:sol-update}
\begin{figure}[h!t]
    \centering
    \includegraphics[trim={0.1in 0.75in 0.1in 0.75in}, clip,width=0.75\textwidth]{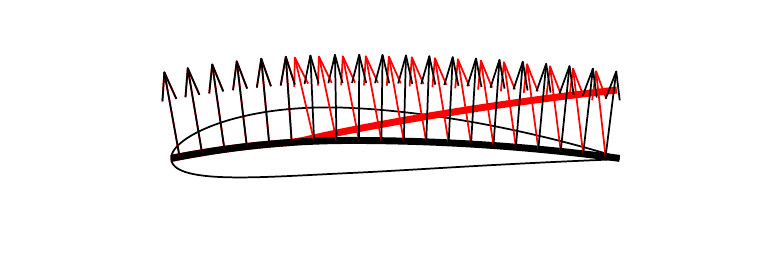}
    \caption{An illustration of the decambering implemented in the VLM by rotating normal vectors in situ. The modified normal vectors (red arrows) are normal to the modified camberline (red line), but located on the original camberline (black line).}
    \label{fig:decam-in-place}
\end{figure}

Once the decambering parameters have been calculated, the shape of the decambering flap is obtained using \eref{eqn:decambering-camberline}.
%
% The slope of the decambering flap is given by
% \begin{align}
%     \diff{z}{x}(x) &= \begin{cases}
%     0 & x < f \\
%     2 A x + B & x \geq f
%     \end{cases}
% \end{align}
%
% Since the geometry is represented by discrete panels, the normal vector of every panel is tilted by the slope at the location of its collocation point ($\left.\diff{z}{x}\right\vert_{i,\text{colloc}}$).
% \begin{align}
%     m &= \left.\diff{z}{x}\right\vert_{i,\text{colloc}} \\
%     \Delta\vec{n}_{i, \text{decambering}} &= \left\{\sin\left(m\right); 0; \cos\left(m\right)-1 \right\}
% \end{align}
%
% In case the separation location falls within a panel, its inclination is multiplied by the fraction of the panel that encounters separated flow.
%
As illustrated in \fref{fig:decam-in-place}, the normal vectors are merely rotated in their original locations without moving the collocation points to the location of the modified camberline, since doing so would require an expensive recalculation of the aerodynamic influence coefficients for the VLM.
Thereafter, steps 1--6 are repeated until the solution has converged. At higher angles of attack, large decambering flaps are required to model the significant separated flow over the wing. At these angles, typically $\alpha > 25\degree$, the solution from the previous angle of attack can be used as the initial solution to aid convergence.

 \section{Results}
\label{sec:unswept-results}

The low-order method (LOM) described above was tested for multiple unswept wing geometries using viscous airfoil data, obtained from wind-tunnel experiments and from 2D RANS CFD solutions, as input. The \methodname was tested using airfoils of various maximum thickness for rectangular and tapered wing planforms. In addition to steady flight conditions, the \methodname was tested in the quasi-steady regime by applying a small, constant roll-rate to the wings. Results from the \methodabbr are compared against experimental observations from literature and 3D RANS CFD solutions for the geometries and flight conditions listed in \tref{tab:geoms}

\begin{table}[h!]
\centering
\caption{Summary of cases presented}
\label{tab:geoms}
\renewcommand{\arraystretch}{1.5}
\begin{tabular}{  c  c  c  c  c  }
\hline
\hline
 Case & Airfoil & Aspect Ratio & $Re$ & Notes  \\
 \hline
 A\textsubscript{1} & \multirow{3}{*}{NACA 4418} & 6 & \multirow{3}{*}{$0.75\times10^6$} & \multirow{3}{*}{\shortstack{Experimental verification \\[10pt] Source: Ref. \cite{naik_ostowari_nrel} }}\\
 A\textsubscript{2} &  & 9 &  &  \\
 A\textsubscript{3} & & 12 &  &  \\
 \cline{1-5}
 B\textsubscript{1}  & \multirow{2}{*}{NACA 0012} & 8 & \multirow{2}{*}{$3\times10^6$} & {Symmetric airfoil} \\
 B\textsubscript{2}  &  & 12 & & CFD verification \\
 \cline{1-5}
 C\textsubscript{1} & \multirow{3}{*}{NACA 4415} & 8 & \multirow{3}{*}{$3\times10^6$} & \multirow{3}{*}{\shortstack{Cambered airfoil\\[10pt]CFD verification}} \\
 C\textsubscript{2} &  & 12 & & \\
 C\textsubscript{3} &  & 16 & & \\
 \cline{1-5}
%  D & \multirow{2}{*}{NACA 0015} & 8 & \multirow{2}{*}{--} \\
%  E &  & 12 & \\
%  \cline{1-4}
D\textsubscript{1}  & NACA 4415 & 12 & $3\times10^6$ & Tapered wing ($\lambda = 0.5$) \\
 \cline{1-5}
%  E\textsubscript{1}  & \multirow{3}{*}{NACA 4415} & 8 & $pb/2V = 1.369\times{10}^{-2}$ \\
E\textsubscript{1}  & \multirow{2}{*}{NACA 4415} & 12 & \multirow{2}{*}{$3\times10^6$} & Rolling wing; Rectangular planform\\
E\textsubscript{2}  &  & 12 &  & Rollling wing; Tapered planform, $\lambda = 0.5$\\
 \hline
 \hline
\end{tabular}
\end{table}

% \PHNote{Change this paragraph to reflect the new world order}

{The lifting surface for each geometry was calculated and discretized into a lattice of 20 spanwise and 40 chordwise panels. Since the decambering shape is implemented in the \methodabbr by rotating the normal vectors of the panels, insufficient chordwise discretization can cause problems with convergence. Therefore, a fairly large number of chordwise panels (when compared with traditional VLMs) is used to ensure sufficient sensitivity of the method to the decambering shape. The code, implemented in Python 3.6 and optimized using the NumPy 1.16.2 package compiled with the Intel MKL libraries to perform vectorized linear algebra operations, ran on an Apple MacBook Air (2.2Ghz Dual-Core Intel i7) in 8-12 minutes for all cases.}

\secref{sec:exp-results} presents case A, in which results are obtained from the low-order method using the 2D viscous input curves obtained in the wind tunnel experiments by  Ostowari and Naik \cite{naik_ostowari_nrel}.
These predictions are compared against 3D experimental results from the same source.

Results presented in subsequent sections were obtained using 2D viscous input curves obtained from 2D RANS simulations performed using ANSYS Fluent on the NCSU HPC cluster. Details about the CFD simulations are given in \secref{sec:cfd-meth}. %
%
% Results from the \methodabbr are shown along with the CFD solutions for the viscous airfoil curves used as input data in \secref{sec:airfoil-results}
The total wing loads ($C_L, C_D, C_M$) predicted by the \methodabbr are compared against CFD solutions in \secref{sec:total-loads} for a rectangular wing with a symmetric 12\% thick airfoil (Case B) and a cambered 15\% thick airfoil (Case C).
% \methodabbr results show excellent agreement with CFD results.
\secref{sec:spanwise-loads} compares the low-order predictions of spanwise distributions of $C_l$ and $C_m$, and separation lines for these wings against the CFD solutions.
% \PHNote{separation lines not yet plotted.}
The \methodname accounts for the effects of the separated boundary layer by ``decambering'' the wing sections, effectively changing its shape.
The decambered wing-sections are overlaid on the contour of $V/V_\infty$ in \secref{sec:dec-shape} to illustrate the resemblance of the decambered airfoil shape to the shape of the separated boundary layer.
The effectiveness of the \methodname in predicting the characteristics of tapered wings is demonstrated in \secref{sec:tapered-wings} for a wing with taper ratio $\lambda = 0.5$. \secref{sec:rolling-wings} illustrates the utility of the \methodname in quasi-steady cases, such as when a small, constant roll-rate is present, for both rectangular and tapered wings.
Finally, \secref{sec:limitations} presents results for a swept wing and discusses the limitations in the method and motivates efforts to develop a correction for swept geometries.

%\PHNote{Should we show results for the airfoils? Don't think it's necessary}

\subsection{Experimental Validation}
\label{sec:exp-results}
% \PHNote{Rewrite}
The 2D viscous operating curves used as input to the low-order method can be obtained experimentally or from computational solutions. Here, we present the results obtained from the \methodname for rectangular NACA 4418 wings of three different aspect ratios using input curves obtained from Ostowari ans Naik \cite{naik_ostowari_nrel}.
A key requirement for the nonlinear decambering approach is the knowledge of the separation location. Since the separation curve ($f$ vs. $\alpha$) for the airfoil is not usually easily available from experiments, Beddoes' method (\eref{eqn:beddoes-f}) is used to calculate an approximate separation curve.

% \tref{tab:exp-geoms} lists the wing geometries for which results are presented in this section along with the sources for the input (2D) curves and the 3D experimental comparison data.

% \begin{table}[h!]
%     \centering
%     \caption{Summary of cases presented for experimental verification}
%     \label{tab:exp-geoms}
%     \renewcommand{\arraystretch}{1.5}
%     \begin{tabular}{ | c | m{2in} | c | m{2in} | }
%         \hline
%         \hline
%         Case & Airfoil & Aspect Ratio & Notes \\
%         \hline
%         H & NACA 4418 & ... & 2D source: \cite{naik_ostowari_nrel}; 3D source: \cite{naik_ostowari_nrel}; $Re = 0.75\times10^6$ \\
%         \hline
%         I & Root: NACA 4424; Tip: NACA 4412 & 12.06 &
%         Taper ratio: 0.4; Washout: $3\degree$ ; 2D source: \cite{abbott}; 3D source: \cite{McVeigh1971}; $Re = 2.87\times 10^6$ \\
%         \hline
%     \end{tabular}
% \end{table}

\begin{figure}[!h]
    \centering
    \includegraphics[width=4in]{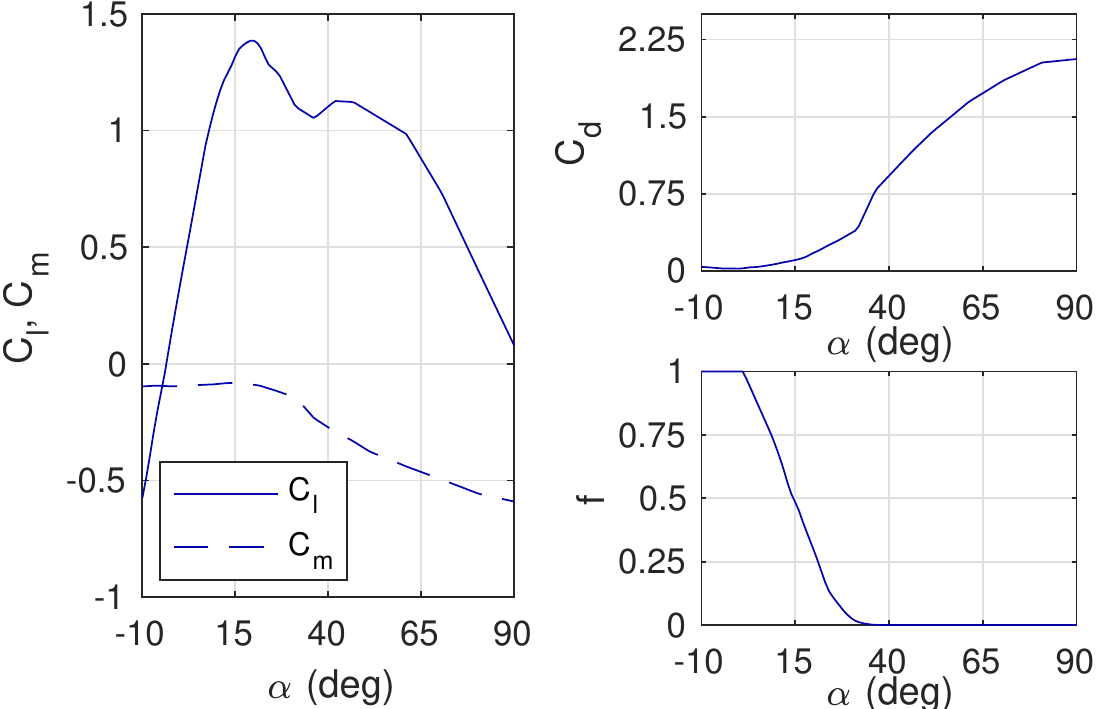}
    \caption{The viscous lift, drag, and moment curves for the NACA 4418 airfoil obtained from experimental tests by Naik and Ostowari \cite{naik_ostowari_nrel}, and the separation curve calculated using Beddoes' model}
    \label{fig:exp-coeffs}
\end{figure}

Wind tunnel results used in Case A are from the experiments \revnote{which were}{\#1.3} performed by Ostowari and Naik \cite{naik_ostowari_nrel} in the Texas A\&M University wind tunnel. Reflection-plane models of various NACA 44XX family wings were used to obtain force and moment curves at angles of attack ranging from $-10\degrees{}$ to $110\degrees{}$.
Data was obtained at a range of Reynolds numbers for wings of aspect ratios 6, 9, and 12, and for a wing spanning the entire test section ($\ar = \infty$). The results shown here use the viscous curves at $Re = 0.75\times10^6$. The separation curve is calculated using \eref{eqn:beddoes-f} and the lift curve for the NACA 4418 airfoil given in Ref. \cite{naik_ostowari_nrel}.
It was seen that at certain angles of attack, \eref{eqn:beddoes-f} gives a value for $f$ that is greater than 1.
In such cases, the value of $f$ is simply set to 1. The experimentally obtained lift, drag, and moment, and calculated separation curves are shown in \fref{fig:exp-coeffs}.

\begin{figure*}[!h]
    \centering
    \includegraphics[width=\figwidth]{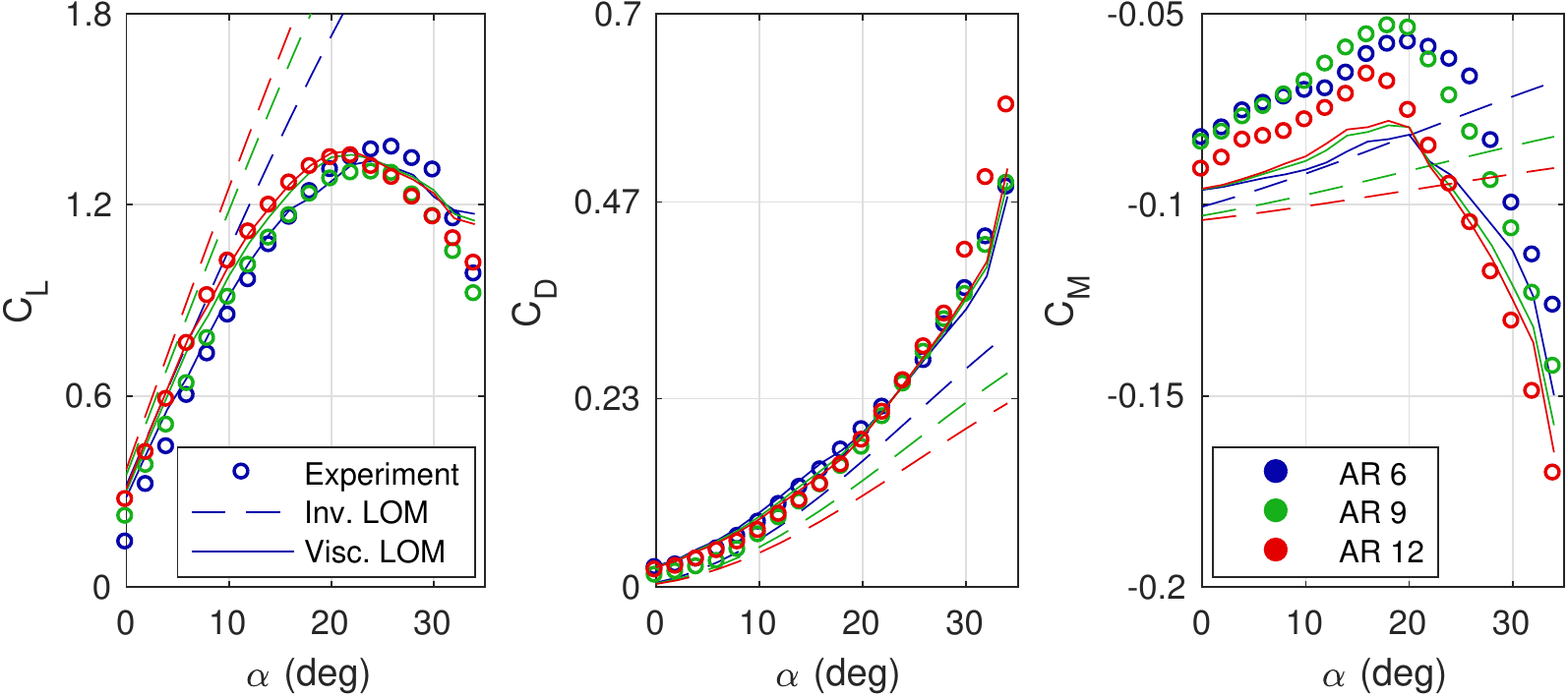}
    \caption{Total coefficients of lift, drag, and pitching moment  for the NACA 4418 wings (Case A\textsubscript{1} -- A\textsubscript{3})}
    \label{fig:exp-n4418-coeffs}
\end{figure*}

Using these input curves, the low-order method can predict the loads on 3D wings of various aspect ratios.
The predictions from the low-order method for the NACA 4418 wings of aspect ratios 6, 9, and 12 are shown in \fref{fig:exp-n4418-coeffs} %
% \begin{figure*}[!h]
%     \centering
%     \begin{subfigure}[t]{\textwidth}
%         \centering
%         \includegraphics[width=\figwidth]{figs/eps_fig/naca4418_ar6_swp0_coeffs.eps}
%         \caption{$\ar{6}$\label{sfig:n4418-ar6-coeffs}}
%     \end{subfigure}%

%     \begin{subfigure}[t]{\textwidth}
%         \centering
%         \includegraphics[width=\figwidth]{figs/eps_fig/naca4418_ar9_swp0_coeffs.eps}
%         \caption{$\ar{9}$\label{sfig:n4418-ar9-coeffs}}
%     \end{subfigure}%

%     \begin{subfigure}[t]{\textwidth}
%         \centering
%         \includegraphics[width=\figwidth]{figs/eps_fig/naca4418_ar12_swp0_coeffs.eps}
%         \caption{$\ar{12}$\label{sfig:n4418-ar12-coeffs}}
%     \end{subfigure}%

%     \caption{Total coefficients of lift, drag, and pitching moment  for the NACA4418 wings (Case J)}
%     \label{fig:exp-n4418-coeffs}
% \end{figure*} %
 %
The low-order predictions for lift and drag show excellent agreement with the experimentally obtained values.
The low-order method correctly predicts the increasing lift-curve slope with increase in aspect ratio at low angles of attack.
As the angle of attack increases, the viscous low-order method correctly predicts stall and the associated drop in $C_L$ and rise in $C_D$.
The maximum $C_L$ and stall angle predictions from the viscous low-order method are within 5\% of the experimental values.
The low-order prediction for moment agrees well with experimental result for the $\ar 12$ case.
However, we see that for the smaller aspect ratios, the low-order moment prediction starts to deviate from the experimental result and the error increases with decreasing aspect ratios. This is thought to be due to the interactions of the detached wingtip vortices with the wing, which become important at low aspect ratios but are not modeled by the current low-order method.

%\PHNote{Residual history?}

\subsection{CFD Methodology for 3D Wings}
\label{sec:cfd-meth}
The low-order method does not require any input data from 3-dimensional CFD solutions. However, 3D CFD solutions for each of the wing geometries described above were obtained to evaluate the accuracy of the low-order method. Using the procedure described in Ref. \cite{Jamwal2018}, body-conforming structured meshes having a wall $y^+ = 1$  were generated for each geometry using the multi-blocking Hexa algorithm in ANSYS ICEM-CFD. These meshes, having cell counts ranging from 20M--45M cells, were used to obtain time-accurate solutions at $Re = 3\times10^6$ in ANSYS Fluent. A physical timestep of 0.01s was used for a total of 300 timesteps. The Spalart-Allmaras model was used for turbulence closure.
\revnote{Time-accurate simulations were performed to ensure that the CFD solutions would converge at the high post-stall angles of attack where the inherently unsteady flow prevents steady simulations from converging. The CFD results shown in this work are the mean values of each quantity over one oscillation.}{\#3.14}
A detailed explanation of the CFD methodology is given in Ref. \cite{AbhimanyuThesis}. The total and spanwise load distributions are obtained from the CFD solutions for comparison with low-order predictions. The separation line is obtained from plots of skin-friction lines on the upper surface of the wing.

\newcommand{\clmax}{\ensuremath{C_{l, \text{max}}}}
\newcommand{\CLmax}{\ensuremath{C_{L, \text{max}}}}

\subsection{Wing Lift, Drag, and Moment Predictions}
\label{sec:total-loads}

\begin{figure}[!h]
    \centering
    \includegraphics[width=4in]{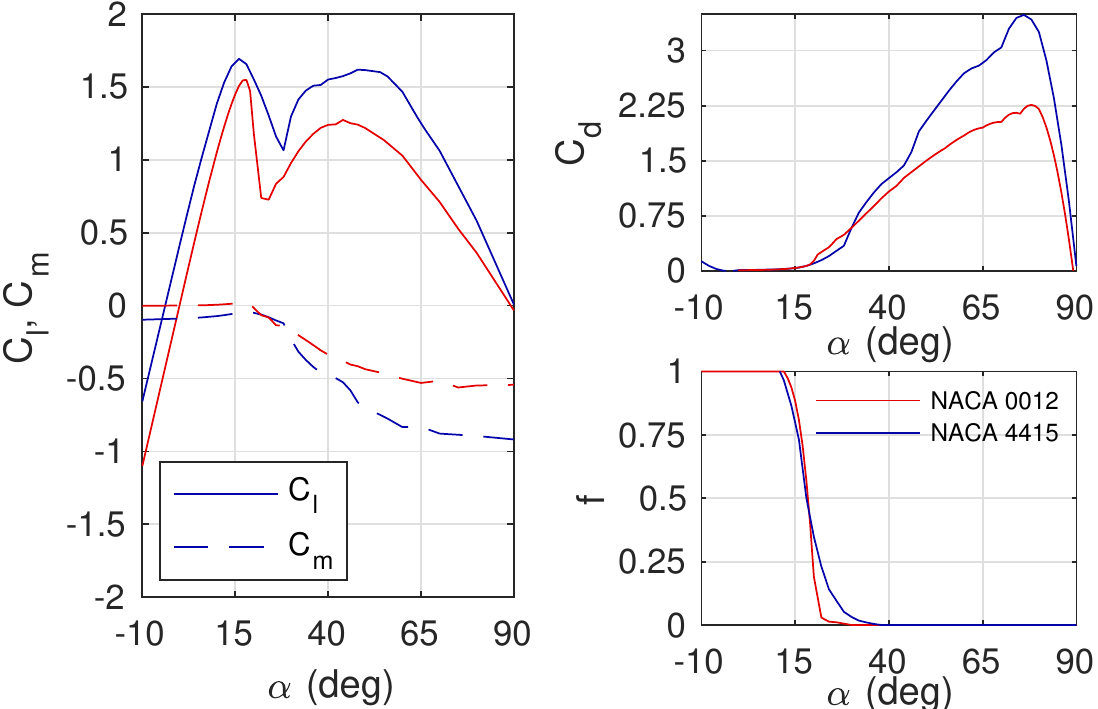}
    \caption{The viscous lift, drag, moment, and separation curves for the NACA 0012 and NACA 4415 airfoils obtained from 2D CFD soluions}
    \label{fig:cfd-coeffs}
\end{figure}

The total coefficients of lift, drag, and pitching moment  up to $\alpha=35\degree$ are presented below. For all these cases, viscous input curves for $C_l$, $C_d$, $C_m$, and $f$ vs. $\alpha$, shown in \fref{fig:cfd-coeffs}, were obtained from 2D CFD solutions at the appropriate Reynolds numbers.

\subsubsection{NACA0012 Wings: Case B}
\begin{figure*}[!h]
        \centering
        \includegraphics[width=\figwidth]{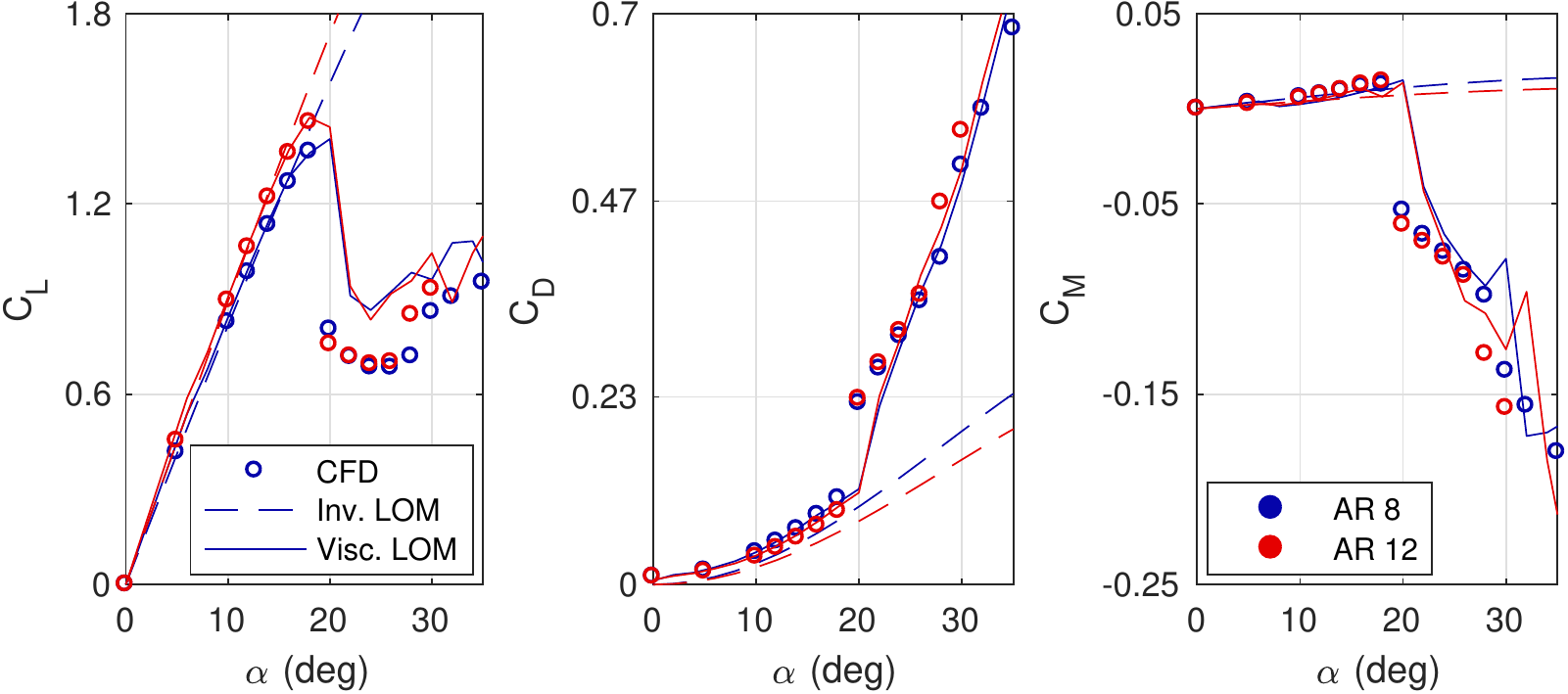}
        \caption{Total coefficients of lift, drag, and pitching moment  for the NACA0012 wings (Cases B\textsubscript{1}--B\textsubscript{2})}
        \label{fig:n0012-coeffs}
    \end{figure*}

The low-order predictions of $C_L, C_D, C_M$ vs. $\alpha$ for the NACA0012 wings are shown in
% \multiref{f}{fig:n0012-ar8-coeffs}{fig:n0012-ar12-coeffs}
\fref{fig:n0012-coeffs}.
At low angles of attack, the inviscid low-order method correctly predicts the loads on the wings, and no additional decambering is required. As the angle of attack increases to $16\degrees{}$ and beyond, the wings begin to stall and the inviscid low-order method does not predict the associated drop in lift and moment, and increase in drag. The viscous \methodname is able to correctly predict these effects of stall.

% \begin{figure*}[!h]
%     \centering
%     \begin{subfigure}[t]{\textwidth}
%         \centering
%         \includegraphics[width=\figwidth]{figs/eps_fig/naca0012_ar8_swp0_coeffs.eps}
%         \caption{$\ar{8}$\label{sfig:n0012-ar8-coeffs}}
%     \end{subfigure}%

%     \begin{subfigure}[t]{\textwidth}
%         \centering
%         \includegraphics[width=\figwidth]{figs/eps_fig/naca0012_ar12_swp0_coeffs.eps}
%         \caption{$\ar{12}$\label{sfig:n0012-ar12-coeffs}}
%     \end{subfigure}%
%     \caption{Total coefficients of lift, drag, and pitching moment  for the NACA0012 wings (Cases D and E)}
%     \label{fig:n0012-coeffs}
% \end{figure*}

\subsubsection{NACA4415 Wings: Case C}

\begin{figure}[!h]
    \centering
    \includegraphics[width=\figwidth]{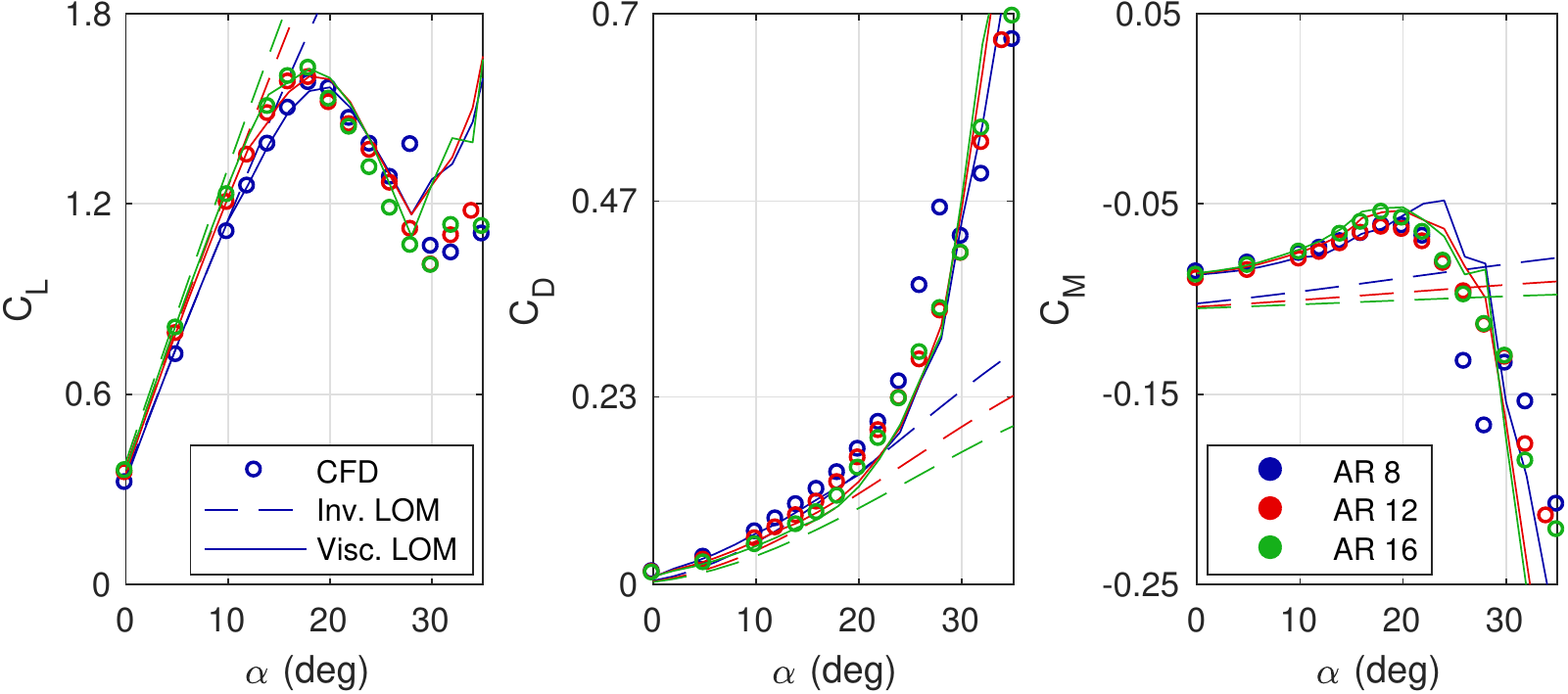}
    \caption{Total coefficients of lift, drag, and pitching moment  for the NACA4415 wings (Cases C\textsubscript{1}--C\textsubscript{3})}
    \label{fig:n4415-coeffs}
\end{figure}

\fref{fig:n4415-coeffs}
% \multiref{f}{fig:n4415-ar8-coeffs}{fig:n4415-ar16-coeffs}
shows the variation of $C_L, C_D$, and  $C_M$ vs. $\alpha$ for the NACA4415 wings.
At low angles of attack, the lift predictions from the inviscid and viscous low-order methods agree well with CFD results.
As the angle of attack increases, the inviscid method does not model the effects of flow separation, and hence the predicted $C_L$ is unsurprisingly higher than the viscous $C_L$ obtained from CFD solutions.
The $C_L$ results from the viscous LOM, however, match CFD results excellently.
The stall angle and the drop in $C_L$ after stall is predicted well for all aspect ratios.
As the angle of attack is increased further beyond $\alpha = 25\degree$, there is massively separated flow on the upper surface of the wings. The flow at such high angles of attack is inherently unsteady, with large stall cells and leading-edge vortex shedding present in the CFD solutions. The low-order method does not model these phenomena, but the predicted coefficients show acceptable agreement with the CFD solutions.

As seen with the lift comparisons, the drag predictions from both methods match CFD results well at low angles of attack.
Interestingly, the drag for the $\ar{8}$ wing is predicted well by the inviscid and viscous low-order methods even at high $\alpha (\approx 20\degree)$ where significant flow separation exists. This is because induced drag, which is predicted well by the inviscid method, is the major contributor to the total drag for the lower aspect ratios. % where the wingtip vortices affect the flow over a large portion of the wing.
As the aspect ratio increases, the induced drag is supplemented by profile drag. This increase in drag is accurately predicted by the viscous LOM.

There is a significant discrepancy in the prediction from the inviscid method for pitching moment even at low $\alpha$. %\PHNote{explanation?}.
This discrepancy is rectified by the decambering method, and the viscous LOM prediction agrees well with CFD. As the angle of attack increases, the viscous LOM accurately predicts the moment break and the angle at which this occurs.
Comparing the results for the three aspect ratios, it can be observed that the viscous LOM predictions generally improve as the aspect ratio increases. This trend occurs because the behavior of the higher $\ar{}$ wings is closer to that of the airfoil.

\subsection{Comparison of Spanwise Distributions of Lift and Moment}
\label{sec:spanwise-loads}

\begin{figure*}
    \centering
    \begin{subfigure}[t]{0.45\textwidth}
        \centering
        \includegraphics[width=0.9\textwidth]{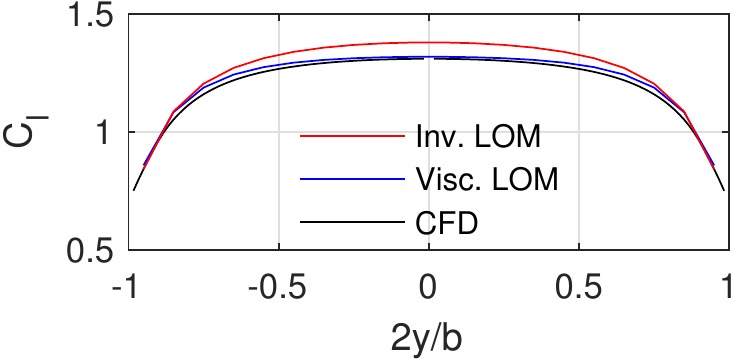}
        \caption{$\alpha=10\degree$\label{sfig:4415-ar12-cldist-a10}}
    \end{subfigure}%
    ~
    \begin{subfigure}[t]{0.45\textwidth}
        \centering
        \includegraphics[width=0.9\textwidth]{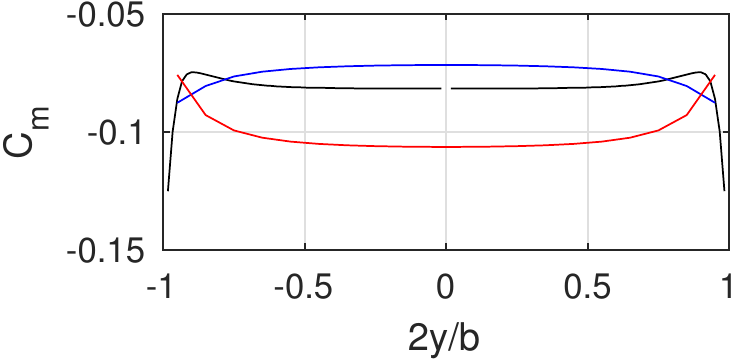}
        \caption{$\alpha=10\degree$\label{sfig:4415-ar12-cmdist-a10}}
    \end{subfigure}%

    \begin{subfigure}[t]{0.45\textwidth}
        \centering
        \includegraphics[width=0.9\textwidth]{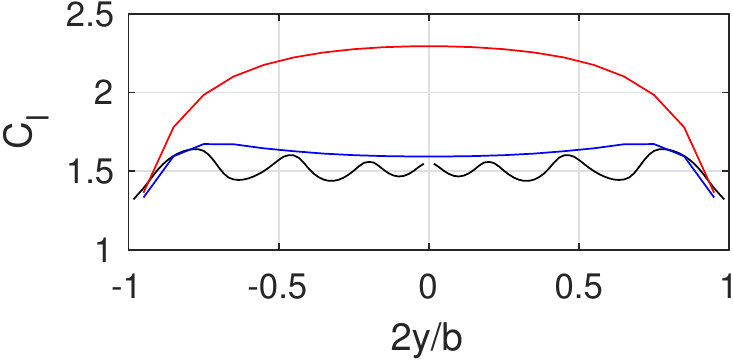}
        \caption{$\alpha=20\degree$\label{sfig:4415-ar12-cldist-a20}}
    \end{subfigure}%
    ~
    \begin{subfigure}[t]{0.45\textwidth}
        \centering
        \includegraphics[width=0.9\textwidth]{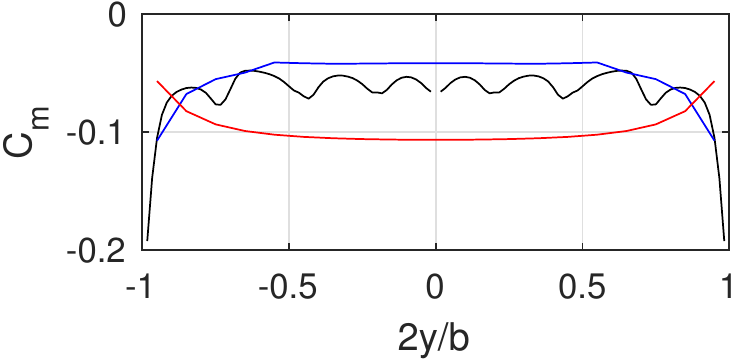}
        \caption{$\alpha=20\degree$\label{sfig:4415-ar12-cmdist-a20}}
    \end{subfigure}%

    \begin{subfigure}[t]{0.45\textwidth}
        \centering
        \includegraphics[width=0.9\textwidth]{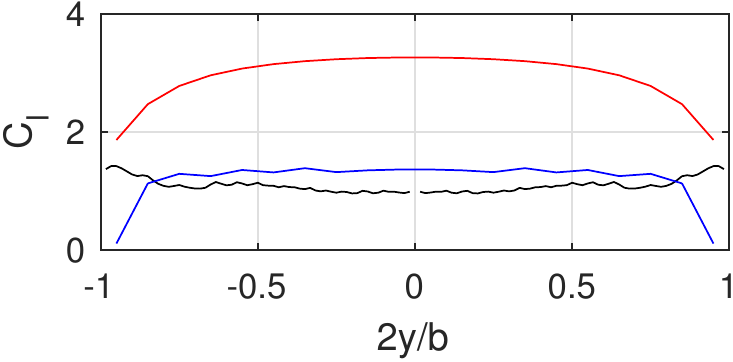}
        \caption{$\alpha=32\degree$\label{sfig:4415-ar12-cldist-a32}}
    \end{subfigure}%
    ~
    \begin{subfigure}[t]{0.45\textwidth}
        \centering
        \includegraphics[width=0.9\textwidth]{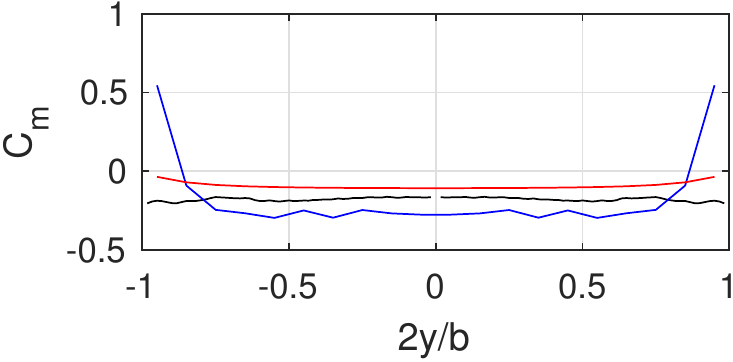}
        \caption{$\alpha=32\degree$\label{sfig:4415-ar12-cmdist-a32}}
    \end{subfigure}%

    \caption{Spanwise distributions of $C_l$ and $C_m$ at pre- and post-stall angles of attack from CFD (black), inviscid LOM (red), and viscous LOM (blue) for the NACA4415 $\ar{12}$ wing (Case C\textsubscript{2})}
    \label{fig:n4415-ar12-cldist}
\end{figure*}

\fref{fig:n4415-ar12-cldist} shows the spanwise distributions of section lift coefficient ($C_l$) and pitching moment  ($C_m$) for the NACA 4415 $\ar{12}$ wing at angles of attack before stall ($\alpha=10\degree$), slightly post-stall ($\alpha=20\degree$), and well beyond stall ($\alpha=32\degree$) at which the boundary layer separates close to the leading edge on a large portion of the wing.
Before stall, the inviscid prediction is close to the viscous solution from CFD. As the angle of attack is increased, the wing stalls at the root. The drop in lift and moment on the inboard sections is predicted well by the viscous LOM. Well beyond stall, the low-order prediction agrees quite well with the CFD solution.
%The accuracy of the low-order prediction at the wingtips is affected by the wingtip vortices. \PHNote{this sentence...}
The large discrepancy between CFD and the low-order method at the wingtips is attributed to the absence of a model in the low-order method to predict the detachment of the tip vortices at higher angles of attack. This discrepancy becomes more evident for the lower aspect-ratio wings, for which the wingtip vortices affect flow over a considerable portion of the wing.

\begin{figure}
    \centering
    \includegraphics[width=3in]{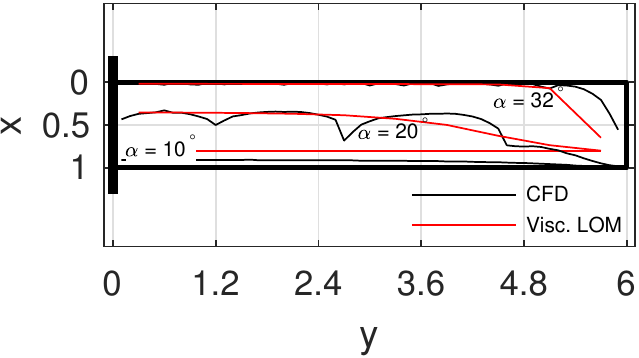}
    \caption{Separation line predicted by the LOM (blue) and CFD (black) for angles of attack ranging from pre-stall to post-stall (Case C\textsubscript{2}). Right half of the wing is shown.}
    \label{fig:naca4415_ar12_swp0_allalpha_fdist}
\end{figure}

\fref{fig:naca4415_ar12_swp0_allalpha_fdist} compares the separation line predicted by the \methodabbr with CFD predictions at the same angles of attack as above. At $\alpha=10\degree$ (well before stall), there is only a small amount of separation, indicated by the separation line being close to $f=1$ (trailing-edge) at all sections. The decambering method models the drop in lift using a trailing-edge flap hinged at the separation line. The other decambering parameters at a section ($\delta_l, m$) depend on the value of $f$. If $f \approx 1$, unphysically large values of $\delta_l$ and $m$ are required to model even a small drop in lift. Therefore, in the low-order method, the aft-most location of the hinge for the decambering flap is constrained to $f \leq 0.8$. As the angle of attack increases beyond stall ($20\degree$), we see significant separation over large portions of the wing. The undulations in the separation line from CFD indicate the presence of stall cells on the upper surface of the wing. While these stall cells are not predicted by the low-order method, the overall agreement of the predicted separation line with the CFD solution is remarkably good. As the angle of attack is increased well beyond stall ($\alpha=32\degree$), we see from the CFD solution that most of the wing experiences fully separated flow. This separation-line behavior is again predicted well by the low-order method. %
%
% \PHNote{What about other spanwise distributions? Should I just refer to my dissertation?}
Spanwise $C_l$, $C_m$, and $f$ distributions for the other geometries given in \tref{tab:geoms} agree similarly well with CFD solutions and are included in Ref. \cite{PranavThesis}.

\subsection{Comparison of Decambering Shape with CFD Velocity Contours}
\label{sec:dec-shape}

As the angle of attack increases, the separated boundary layer changes the effective shape of the wing.
The nonlinear decambering method models the effects of a separated boundary layer using a parabolic decambering flap to simultaneously achieve a drop in both lift and moment.
% \multiref{f}{fig:n4415-ar12-y1.5-blpics}{fig:n4415-ar12-y5.1-blpics}
% This observation is applied later in \cref{ch:wakepred} to estimate the location of the wake behind the wing, which is used by the \methodabbr to predict deep stall of downstream surfaces.
%
% \PHNote{Is this size good? or too small?}
\begin{figure}[!ht]
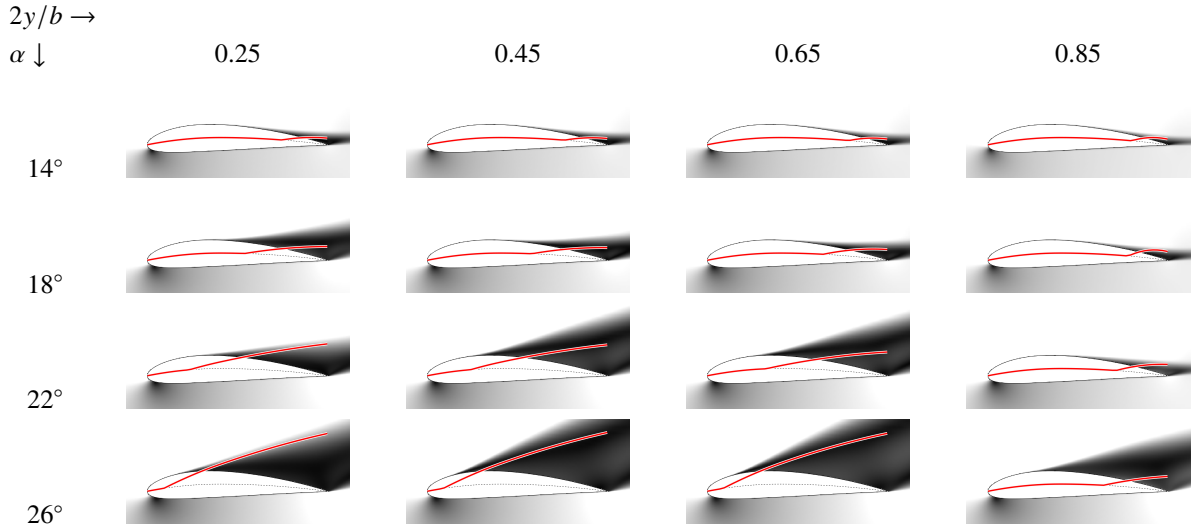

    \centering
    \begin{tabular}{p{0.38in} c c c c}
        %\hline
        %\hline
        %\diagbox[innerwidth=0.35in]{$\alpha$}{$2y/b$}
        $2y/b\rightarrow$ \\ $\alpha\downarrow$ & 0.25  & 0.45 & 0.65 & 0.85 \\
        %\hline
        \hfil$14\degree$ & \input{blplots/a1-y1.tex} & \input{blplots/a1-y2.tex} & \input{blplots/a1-y3.tex} & \input{blplots/a1-y4.tex} \\
        \hfil$18\degree$ & \input{blplots/a2-y1.tex} & \input{blplots/a2-y2.tex} & \input{blplots/a2-y3.tex} & \input{blplots/a2-y4.tex} \\
        \hfil$22\degree$ & \input{blplots/a3-y1.tex} & \input{blplots/a3-y2.tex} & \input{blplots/a3-y3.tex} & \input{blplots/a3-y4.tex} \\
        \hfil$26\degree$ & \input{blplots/a4-y1.tex} & \input{blplots/a4-y2.tex} & \input{blplots/a4-y3.tex} & \input{blplots/a4-y4.tex} \\
        %\hline
        %\hline
    \end{tabular}
    \caption{The decambered camberline (red) at spanwise stations of the NACA 4415 $\ar{12}$ wing (Case C\textsubscript{2}) overlaid on a contour plot of the velocity magnitude $V/V_\infty$ from pre-stall to post-stall angles of attack}
    \label{fig:n4415-blpics}
\end{figure}%

\fref{fig:n4415-blpics} compares the geometry of the decambered wing at multiple sections with contour plots showing the ratio of velocity magnitude to the freestream velocity ($V/V_\infty$). At $\alpha=14\degree$, the flow is mostly attached at all sections of the wing.
A small decambering flap is sufficient to accurately model the effective shape change due to the boundary layer. As the angle of attack is increased to $18\degree$, the separation point moves closer to the leading edge, the wing stalls and the boundary layer becomes thicker. The forward movement of the separation point is predicted well by the \methodname at all sections away from the wingtip. The thicker boundary layer is mimicked well by the decambering flap having a larger deflection and trailing-edge height. We also see that the \methodname accurately predicts the tendency of a rectangular wing to stall to the root. Upon increasing the angle of attack further to $22\degree$ and then to $26\degree$, we observe that the separation point at most sections is very close to the leading edge. The decambering flap approximately models the centerline of the thick boundary layer.
This observation was applied in Ref.~\cite{Hosangadi2018} to predict the location of the viscous wake behind the wing, and the velocity profile in the wake without the need for expensive boundary layer calculations. %
% \PHNote{Refer to dissertation?}
Figures showing the comparison of the decambered camberlines for other geometries are included in Ref. \cite{PranavThesis}
\subsection{Predictions for Tapered Wing (Case D)}
\label{sec:tapered-wings}
\begin{figure}[!h]
    \centering
    % \begin{subfigure}[t]{\textwidth}
    %     \centering
        \includegraphics[width=\figwidth]{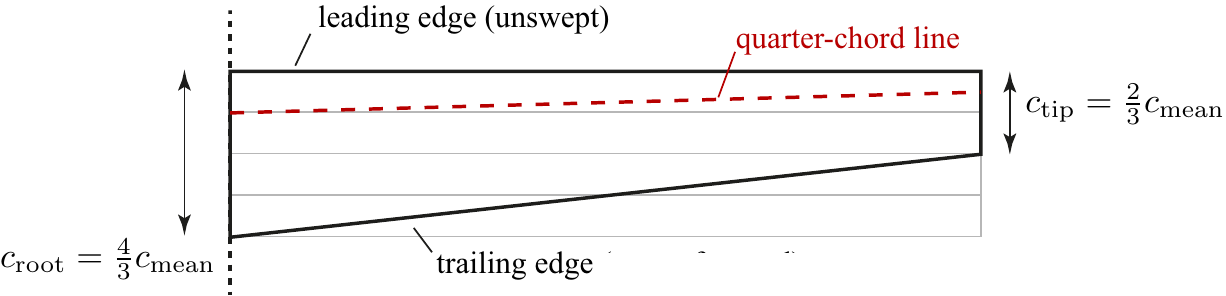}
    %     \caption{\label{sfig:letap-planform}}
    % \end{subfigure}%

    % \begin{subfigure}[t]{\textwidth}
    %     \centering
    %     \includegraphics[width=\figwidth]{figs/eps_fig/qc-unswept-planform.eps}
    %     \caption{\label{sfig:qctap-planform}}
    % \end{subfigure}%
    % \caption{Planform view of the tapered wings with (a) unswept leading edge and (b) unswept quarter-chord line}
    \caption{Planform view of the tapered wing with  unswept leading edge}
    \label{fig:tapered-planforms}
\end{figure}

\begin{figure*}[!h]
    \centering
    % \begin{subfigure}[t]{\textwidth}
    %     \centering
        \includegraphics[width=\figwidth]{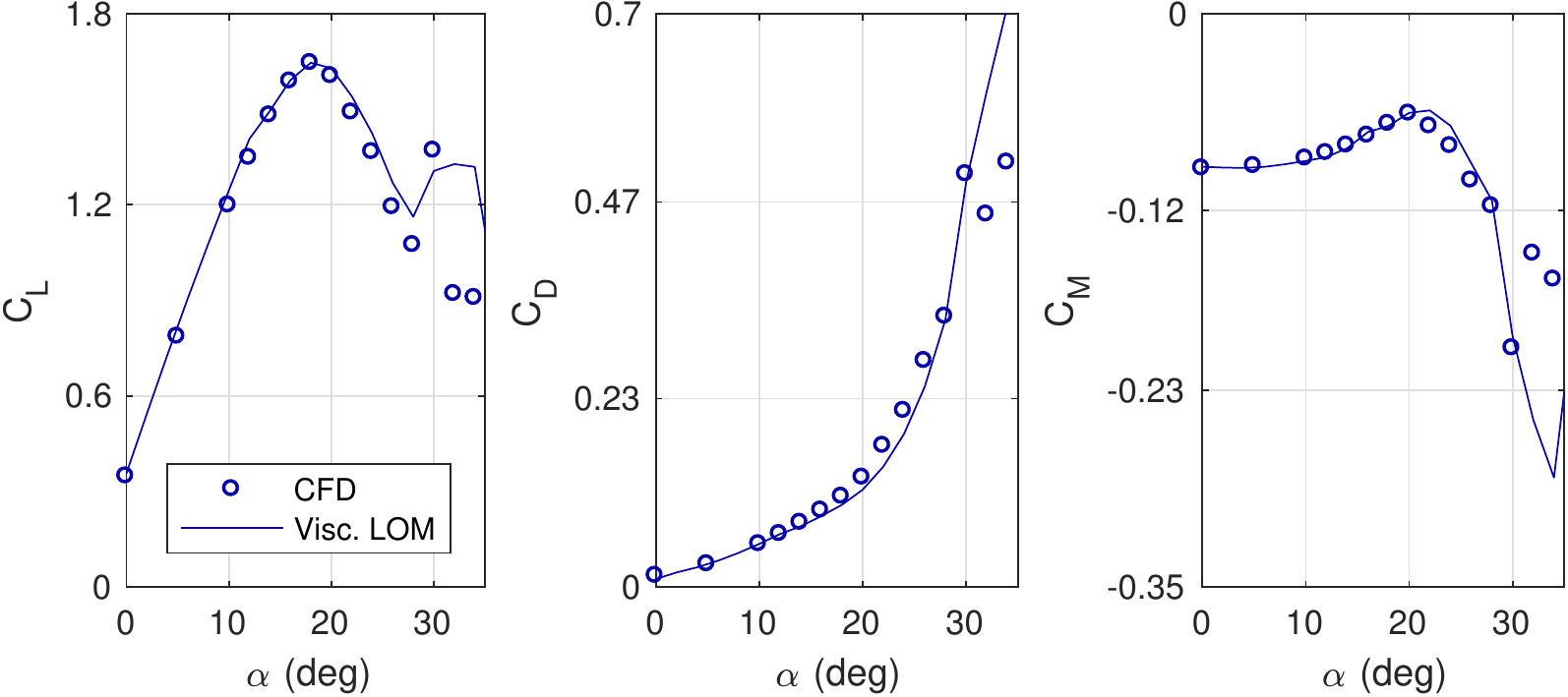}
    %     \caption{Unswept leading edge\label{sfig:4415-ar12-letap}}
    % \end{subfigure}%

    % \begin{subfigure}[t]{\textwidth}
    %     \centering
    %     \includegraphics[width=\figwidth]{{figs/eps_fig/naca4415_taper0.5_qc-unswept_coeffs}.eps}
    %     \caption{Unswept Quarter-chord line\label{sfig:4415-ar12-qctap}}
    % \end{subfigure}%
    % \caption{Total lift, drag, and pitching moment vs. $\alpha$ for the tapered wings from CFD (black) and viscous LOM (blue)}
    \caption{Total lift, drag, and pitching moment vs. $\alpha$ for the tapered wing (Case D) from CFD (black) and viscous LOM (blue)}
    \label{fig:n4415-tap-coeff}
\end{figure*}

\fref{fig:n4415-tap-coeff} shows the coefficients of lift, drag, and pitching moment about root-quarter-chord for the tapered wing (Case F in \tref{tab:geoms}). This wing has an aspect ratio $\ar{}=12$, a taper ratio $\lambda = 0.5$, an unswept leading edge, and a root chord $4/3$ times the mean chord. The planform of the wing is shown in \fref{fig:tapered-planforms}.
% \PHNote{CFD and LOM moments are about (0.25,0,0) not root-quarter-chord}

Viscous low-order results for the tapered wing were obtained using the same 2D viscous curves at each section, i.e. no Reynolds number adjustments were made. The results from the \methodname are shown in \fref{fig:n4415-tap-coeff}.
The viscous LOM accurately predicts the stall angle and the associated drop in lift and moment for the tapered wing.

\begin{figure*}[!h]
    \centering
    \begin{subfigure}[t]{0.45\textwidth}
        \centering
        \includegraphics[width=0.9\textwidth]{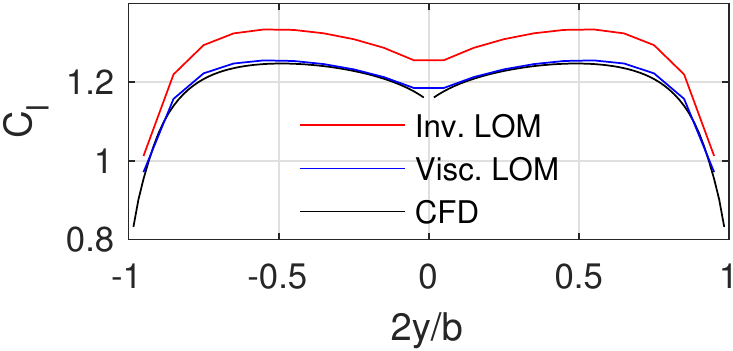}
        \caption{$\alpha=10\degree$\label{sfig:4415-ar12-le-unswept-cldist-a10}}
    \end{subfigure}%
    ~
    \begin{subfigure}[t]{0.45\textwidth}
        \centering
        \includegraphics[width=0.9\textwidth]{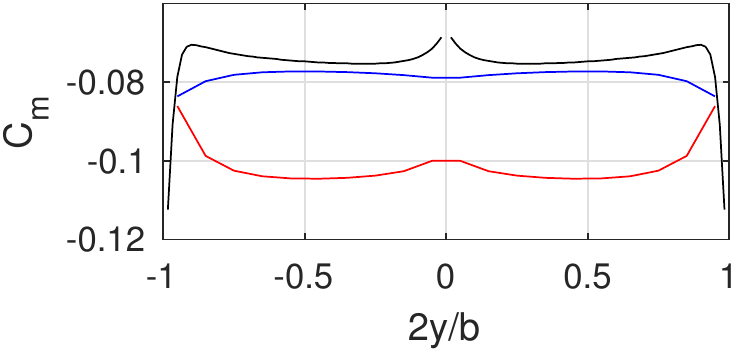}
        \caption{$\alpha=10\degree$\label{sfig:4415-ar12-le-unswept-cmdist-a10}}
    \end{subfigure}%

    \begin{subfigure}[t]{0.45\textwidth}
        \centering
        \includegraphics[width=0.9\textwidth]{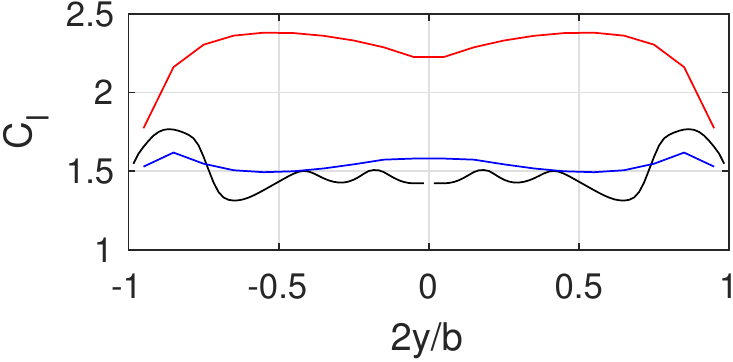}
        \caption{$\alpha=20\degree$\label{sfig:4415-ar12-le-unswept-cldist-a22}}
    \end{subfigure}%
    ~
    \begin{subfigure}[t]{0.45\textwidth}
        \centering
        \includegraphics[width=0.9\textwidth]{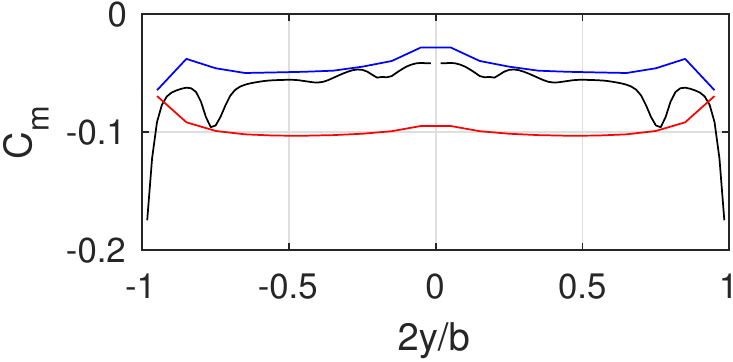}
        \caption{$\alpha=22\degree$\label{sfig:4415-ar12-le-unswept-cmdist-a22}}
    \end{subfigure}%

    \begin{subfigure}[t]{0.45\textwidth}
        \centering
        \includegraphics[width=0.9\textwidth]{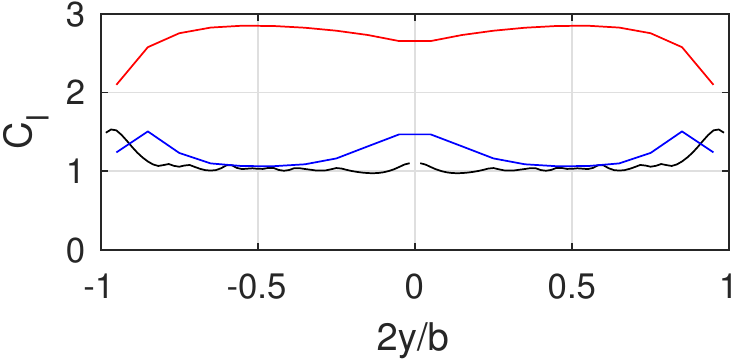}
        \caption{$\alpha=28\degree$\label{sfig:4415-ar12-le-unswept-cldist-a20}}
    \end{subfigure}%
    ~
    \begin{subfigure}[t]{0.45\textwidth}
        \centering
        \includegraphics[width=0.9\textwidth]{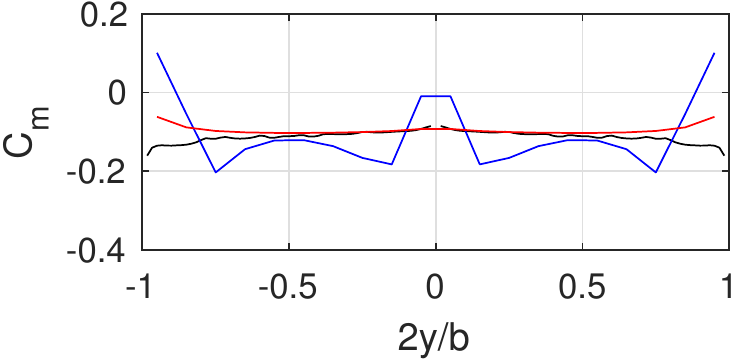}
        \caption{$\alpha=28\degree$\label{sfig:4415-ar12-le-unswept-cmdist-a28}}
    \end{subfigure}%

    \caption{Spanwise distributions of $C_l$ and $C_m$ at pre- and post-stall angles of attack from CFD (black), inviscid LOM (red), and viscous LOM (blue) for the tapered NACA4415 $\ar{12}$ wing (Case D)}
    \label{fig:n4415-ar12-le-unswept-cldist}
\end{figure*}

\begin{figure}[!h]
    \centering
    \includegraphics[width=3in]{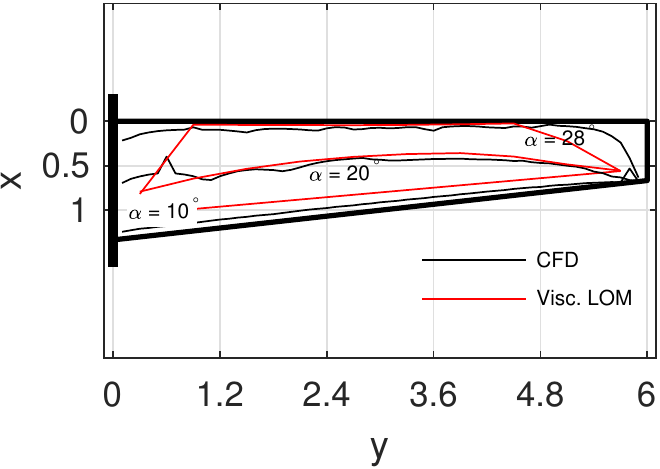}
    \caption{Separation line predicted by the LOM (blue) and CFD (black) for angles of attack ranging from pre-stall to post-stall (Case D).}
    \label{fig:naca4415_ar12_le-unswept_allalpha_fdist}
\end{figure}

From \fref{fig:n4415-ar12-le-unswept-cldist}, it can be seen that the viscous low-order method correctly predicts the spanwise distributions of lift and moment at the pre-stall angle of attack ($\alpha = 10\degree$). %
% The magnitudes of lift and nose-down pitching moment are slightly overpredicted throughout the span of the tapered wing.
As the angle of attack increases to stall and beyond ($ \alpha \geq 20\degree$), the drop in lift and moment is correctly modeled by the decambering approach implemented in the viscous low-order method. It is worth noting that the viscous LOM achieves this using viscous input curves obtained at a single Reynolds number ($Re = 3\times10^6$). The separation line is also accurately predicted by the low-order method, as shown in \fref{fig:naca4415_ar12_le-unswept_allalpha_fdist}.

% \PHNote{Spanwise distributions for tapered wings}

\subsection{Predictions for Wings Undergoing Rolling Motion (Case E)}
\label{sec:rolling-wings}
Although the underlying vortex lattice method used in this work is a steady code, it is capable of making predictions for quasi-steady flow states, such as for wings undergoing small, constant rates of rotation.
These quasi-steady conditions are typical of those experienced by general aviation and transport aircraft.
% In these quasi-steady states, flow features such as leading-edge vortex shedding which are usually observed in highly unsteady flows, are absent.
The examples presented in this section demonstrate the ability of the viscous LOM to predict the variation of total wing lift, drag, and moment coefficients, and their spanwise distributions, for two wings having a roll rate of $\SI{0.1}{\radian\per\second}$ or $\SI{5.73}{\deg\per\second}$ about the chordwise axis. This roll rate causes the left wing ($2y/b < 0$) to move downwards and see an increased effective angle of attack, while the right wing ($2y/b > 0$) moves upwards and experiences a reduced effective angle of attack. Two geometries, each of aspect ratio 12, are presented in this section: one rectangular and one tapered with a taper ratio $\lambda = 0.5$.
To verify the results from the low-order method, CFD solutions were obtained using ANSYS Fluent at select angles of attack before (0\degree, 10\degree), close to (15\degree, 18\degree), and after (20\degree) stall.

\begin{figure*}
    \centering
        \includegraphics[width=\figwidth]{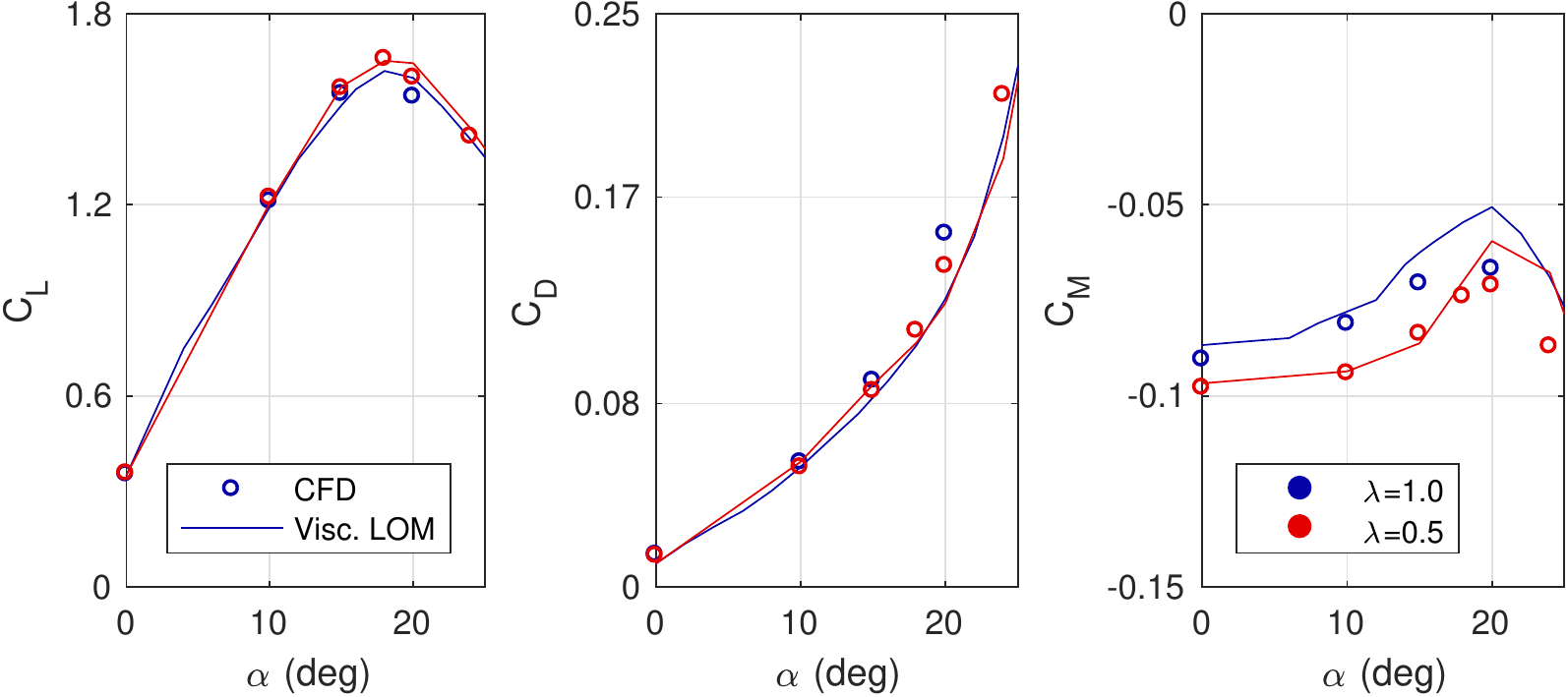}
    \caption{Total coefficients of lift, drag, and pitching moment for the NACA4415 $\ar{12}$ wings (Cases E\textsubscript{1}--E\textsubscript{2}) experiencing a $0.1\si{\radian\per\second}$ roll-rate}
    \label{fig:n4415-rot-coeffs}
\end{figure*}

\begin{figure*}
    \centering
    \begin{subfigure}[t]{0.45\textwidth}
        \centering
        \includegraphics[width=0.9\textwidth]{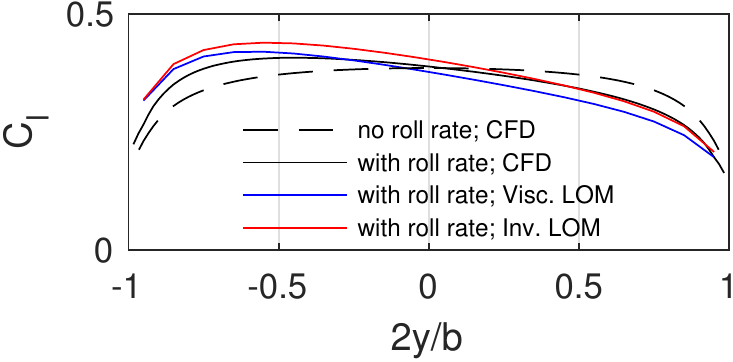}
        \caption{$\alpha=0\degree$\label{sfig:4415-ar12-cldist-a0-rot}}
    \end{subfigure}%
    % ~
    % \begin{subfigure}[t]{0.45\textwidth}
    %     \centering
    %     \includegraphics[width=0.9\textwidth]{{figs/eps_fig/spanwise-dist/naca4415_omega0.1_ar12_swp0_a0.0_cmdist}.eps}
    %     \caption{$\alpha=0\degree$\label{sfig:4415-ar12-cmdist-a0-rot}}
    % \end{subfigure}%

    \begin{subfigure}[t]{0.45\textwidth}
        \centering
        \includegraphics[width=0.9\textwidth]{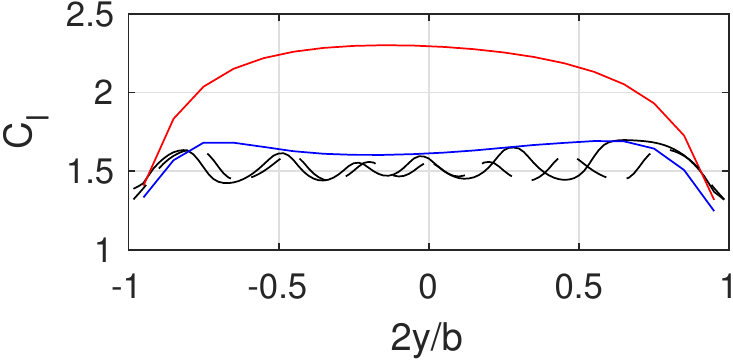}
        \caption{$\alpha=20\degree$\label{sfig:4415-ar12-cldist-a20-rot}}
    \end{subfigure}%
    % ~
    % \begin{subfigure}[t]{0.45\textwidth}
    %     \centering
    %     \includegraphics[width=0.9\textwidth]{{figs/eps_fig/spanwise-dist/naca4415_omega0.1_ar12_swp0_a20.0_cmdist}.eps}
    %     \caption{$\alpha=20\degree$\label{sfig:4415-ar12-cmdist-a20-rot}}
    % \end{subfigure}%

    \caption{Spanwise distributions of $C_l$ at pre- and post-stall angles of attack from CFD (black), inviscid LOM (red), and viscous LOM (blue) for the NACA4415 rectangular $\ar{12}$ wing (Case E\textsubscript{1}) with a $0.1\si{\radian\per\second}$ roll rate}
    \label{fig:n4415-ar12-dist-rot}
\end{figure*}

\begin{figure*}
    \centering
    \begin{subfigure}[t]{0.45\textwidth}
        \centering
        \includegraphics[width=0.9\textwidth]{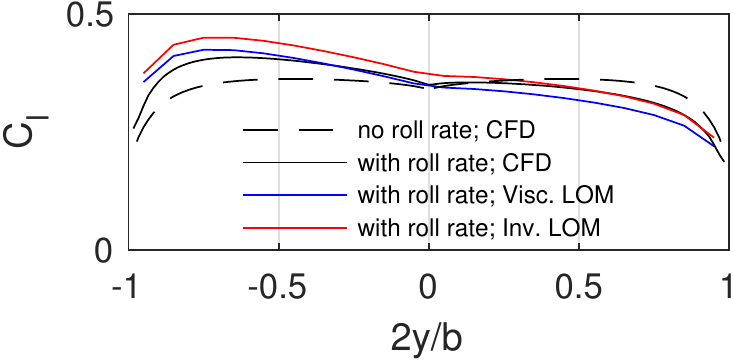}
        \caption{$\alpha=0\degree$\label{sfig:4415-ar12-le-uns-cldist-a0-rot}}
    \end{subfigure}%
    % ~
    % \begin{subfigure}[t]{0.45\textwidth}
    %     \centering
    %     \includegraphics[width=0.9\textwidth]{{figs/eps_fig/spanwise-dist/naca4415_ar12_le-unswept_omega0.1_a0.0_cmdist}.eps}
    %     \caption{$\alpha=0\degree$\label{sfig:4415-ar12-le-uns-cmdist-a0-rot}}
    % \end{subfigure}%

    \begin{subfigure}[t]{0.45\textwidth}
        \centering
        \includegraphics[width=0.9\textwidth]{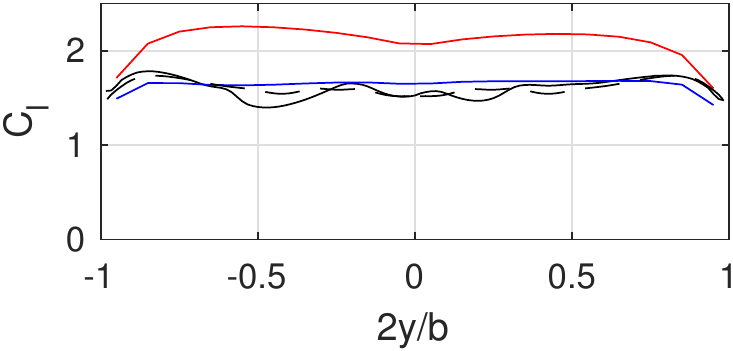}
        \caption{$\alpha=20\degree$\label{sfig:4415-ar12-le-uns-cldist-a20-rot}}
    \end{subfigure}%
    % ~
    % \begin{subfigure}[t]{0.45\textwidth}
    %     \centering
    %     \includegraphics[width=0.9\textwidth]{{figs/eps_fig/spanwise-dist/naca4415_ar12_le-unswept_omega0.1_a20.0_cmdist}.eps}
    %     \caption{$\alpha=20\degree$\label{sfig:4415-ar12-le-uns-cmdist-a20-rot}}
    % \end{subfigure}%

    \caption{Spanwise distributions of $C_l$ at pre- and post-stall angles of attack from CFD (black), inviscid LOM (red), and viscous LOM (blue) for the NACA4415 tapered wing (Case E\textsubscript{2}) with a $0.1\si{\radian\per\second}$ roll rate}
    \label{fig:n4415-tap-dist-rot}
\end{figure*}

\begin{figure*}
    \centering
        \includegraphics{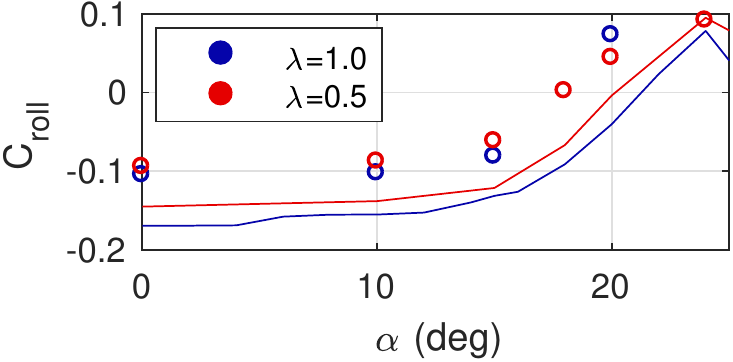}
    \caption{Total rolling moment vs. $\alpha$ on the rectangular and tapered $\ar{12}$ wings from CFD (symbols) and viscous LOM (lines)}
    \label{fig:n4415-croll}
\end{figure*}

% \multiref{f}{fig:rot-ar8-coeffs}{fig:rot-ar12-coeffs}
\fref{fig:n4415-rot-coeffs}
shows the total wing $C_L$, $C_D$, and $C_M$ vs. $\alpha$ variation predicted by the \methodname. The lift and drag predictions from the viscous LOM agree well with CFD solutions. Predictions for pitching moment are seen to deviate from CFD solutions at higher angles of attack.
\multiref{f}{fig:n4415-ar12-dist-rot}{fig:n4415-tap-dist-rot} show the comparison of the spanwise distribution of $C_l$.
For reference, the spanwise distribution of $C_l$ for the wings without any rotational velocity from CFD is plotted using the dashed black line.
At a low angle of attack (0\degree) shown in Figures \ref{sfig:4415-ar12-cldist-a0-rot} and \ref{sfig:4415-ar12-le-uns-cldist-a0-rot}, the rolling motion of the wing causes an increase in lift on the left side (rolling downwards) and a drop in lift on the right side (rolling upwards).
It is this increase in lift on the descending wing that results in roll damping at unstalled conditions.
At $\alpha=20\degree$, this effect is reversed. The lift produced on the left side of the wing is reduced, whereas the right side of the wing produces more lift than the case without any rotation, as seen from Figures \ref{sfig:4415-ar12-cldist-a20-rot} and \ref{sfig:4415-ar12-le-uns-cldist-a20-rot}.
This post-stall behavior that results in loss of roll damping is captured correctly by the LOM.
The variation of total coefficient of rolling moment with angle of attack for the rectangular and tapered wings is shown in \fref{fig:n4415-croll}.
At low angles of attack, the wing experiences a negative rolling moment, i.e. in the direction opposite to the roll. The negative rolling moment indicates that roll damping is present. As the angle of attack increases, the restoring moment reduces, and after stall, the rolling moment acts in the direction of the rotation. The low-order method correctly predicts the loss of roll damping due to stall.

% \begin{figure*}
%     \centering
%     \begin{subfigure}[t]{0.45\textwidth}
%         \centering
%         \includegraphics[width=0.9\textwidth]{{figs/eps_fig/spanwise-dist/naca4415_omega0.1_ar8_swp0_a0.0_cldist}.eps}
%         \caption{$\alpha=0\degree$\label{sfig:4415-ar8-cldist-a0-rot}}
%     \end{subfigure}%
%     ~
%     \begin{subfigure}[t]{0.45\textwidth}
%         \centering
%         \includegraphics[width=0.9\textwidth]{{figs/eps_fig/spanwise-dist/naca4415_omega0.1_ar8_swp0_a0.0_cmdist}.eps}
%         \caption{$\alpha=0\degree$\label{sfig:4415-ar8-cmdist-a0-rot}}
%     \end{subfigure}%

%     \begin{subfigure}[t]{0.45\textwidth}
%         \centering
%         \includegraphics[width=0.9\textwidth]{{figs/eps_fig/spanwise-dist/naca4415_omega0.1_ar8_swp0_a20.0_cldist}.eps}
%         \caption{$\alpha=20\degree$\label{sfig:4415-ar8-cldist-a20-rot}}
%     \end{subfigure}%
%     ~
%     \begin{subfigure}[t]{0.45\textwidth}
%         \centering
%         \includegraphics[width=0.9\textwidth]{{figs/eps_fig/spanwise-dist/naca4415_omega0.1_ar8_swp0_a20.0_cmdist}.eps}
%         \caption{$\alpha=20\degree$\label{sfig:4415-ar8-cmdist-a20-rot}}
%     \end{subfigure}%

%     \caption{Spanwise distributions of $C_l$ and $C_m$ at pre- and post-stall angles of attack from CFD (black), inviscid LOM (red), and viscous LOM (blue) for the NACA4415 rectangular $\ar{8}$ wing (Case E) with a $0.1\si{\radian\per\second}$ roll rate}
%     \label{fig:n4415-ar8-dist-rot}
% \end{figure*}

% \PHNote{Hmm, no significant difference between rect and tapered from LOM an CFD}

% \clearpage

% \PHNote{Final conclusion}

% At low angles of attack, the wing experiences a negative rolling moment, i.e. in the direction opposite to the roll. As the angle of attack increases, this restoring moment reduces and, after stall, the rolling moment acts in the direction of the rotation. This trend is correctly predicted by the low-order method.

 \section{Limitations and Future Extensions of the Nonlinear Decambering Method}
\label{sec:limitations}
The nonlinear decambering method provides excellent predictions for wings of different airfoils and moderate to high aspect ratios, with and without taper and for wings experiencing small roll rates up to angles of attack well beyond stall. However, the accuracy of predictions from the low-order method suffers beyond $\alpha \approx 35\degree$ and for wings having swept planforms. These limitations are discussed in the following sections.

\subsection{Extremely high angles of attack}
\begin{figure}[!h]
    \centering
    \includegraphics[width=0.3\textwidth]{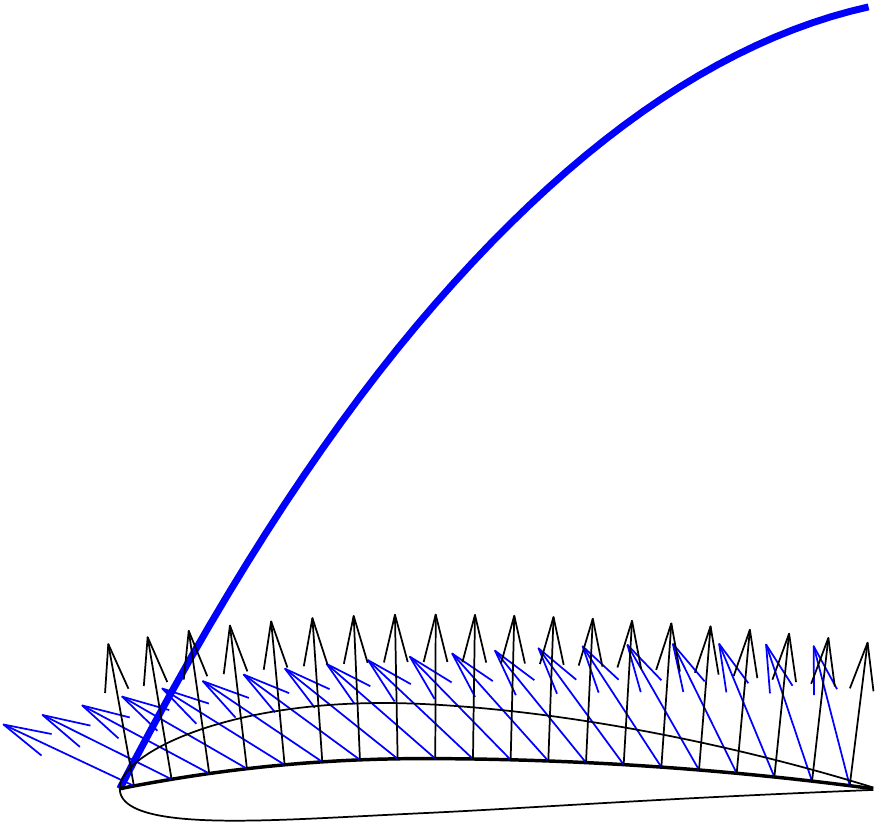}
    \caption{An illustration of the decambering implemented in the VLM by rotating normal vectors in situ at a high $\alpha$. The decambered camberline is no longer adequately approximated by simply rotating the normal vectors.}
    \label{fig:decam-alpha40}
\end{figure}

\newcommand{\bigO}{\ensuremath{\mathcal{O}}}
The decambering method has difficulty converging to a solution beyond $\alpha \approx 35\degree$. This difficulty is a consequence of the way decambering changes the shape of the wing section. As shown in \fref{fig:decam-in-place}, the decambering is applied by simply tilting the normal vectors in place. This approximation is used so that the AIC matrix may be calculated once and then reused for a given geometry so as to speed up the time required to obtain a solution. However, at extremely high angles of attack, such as at $\alpha=40\degree$ shown in \fref{fig:decam-alpha40}, a large decambering flap is required to obtain the required drop in lift and moment. The zero-normal-flow boundary condition is enforced at the collocation points in the direction of the normal vectors. The new camberline is no longer approximated satisfactorily by simply rotating the normal vectors at their original collocation points. However, recalculating the AIC matrix as a result of changing the geometry is a computationally expensive process, requiring $\bigO(n^2)$ computations for a vortex lattice with $n$ ring elements. More research is required to develop a useful compromise between speed and accuracy.

\subsection{Effects of sweep angle}
\begin{figure}[!h]
    \centering
    \includegraphics[width=\figwidth]{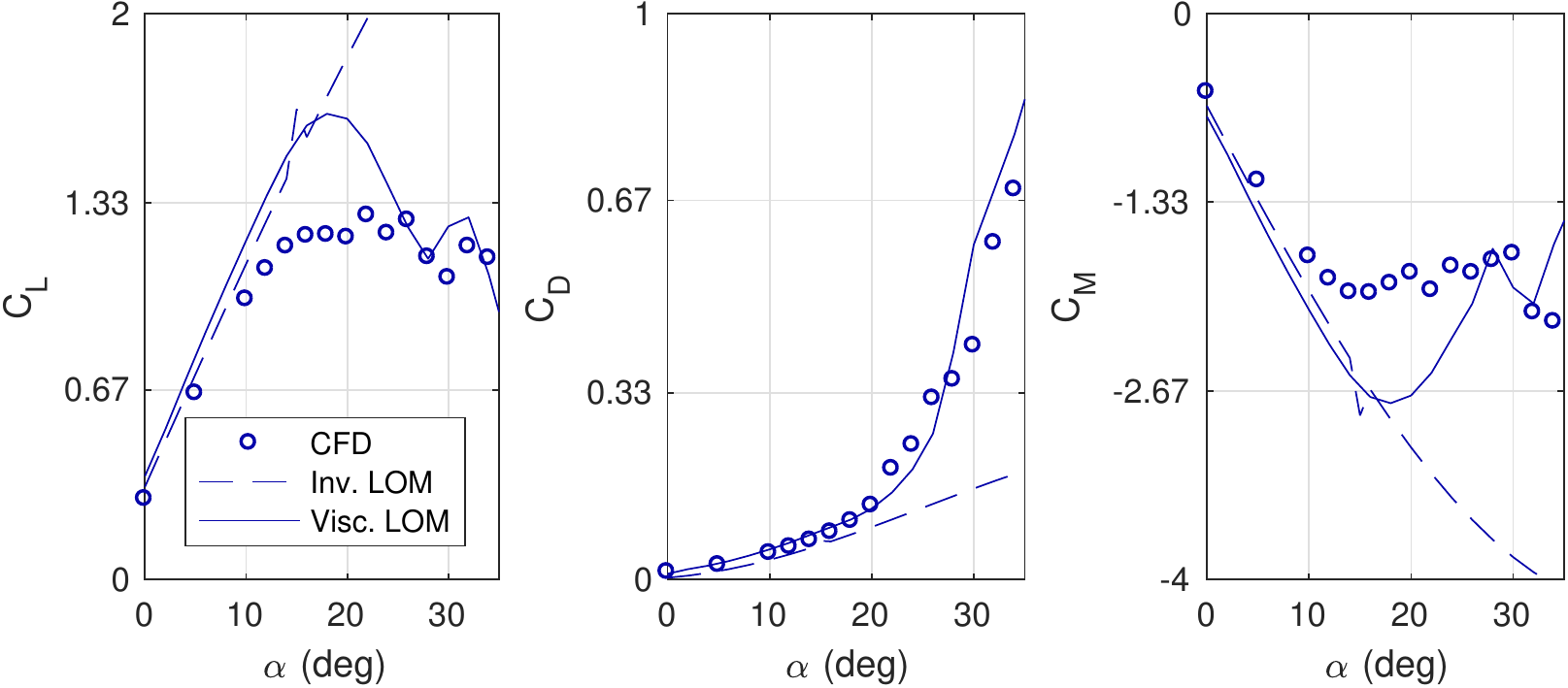}
    \caption{Total lift, drag, and pitching moment vs $\alpha$ for the NACA4415 $\ar{16}$ $30\degree$ swept wing from CFD (black) and the viscous LOM using airfoil input curves (blue)}
    \label{fig:n4415-swpwing-limitation}
\end{figure}

The primary assumption of the decambering method is that the behavior of the sections of the three-dimensional wing is identical to that of the two-dimensional airfoil.
Based on this assumption, the airfoil $C_l$, $C_d$, and $C_m$ vs. $\alpha$ curves are used to obtain the target operating points for each section.
A spanwise pressure gradient exists on swept wings that causes spanwise transport of the separated boundary layer.
This spanwise transport affects the characteristics of the sections of the swept wing and invalidates the assumption that the behavior of each section matches that of the airfoil.
This causes the decambering method to identify the wrong ($C_l, \aeff$) and ($C_m, \aeff$) operating points as the target for the sections, and significantly overpredict $C_L$ and $C_M$ for the swept wing at stall, as shown in \fref{fig:n4415-swpwing-limitation}.
An early approach to correct for these swept-wing effects is discussed in Ref.~\cite{Hosangadi2015}.
% \cref{ch:sweep-effects}.

\subsection{Limitations of RANS CFD}
The limitations of RANS CFD models in predicting highly separated flows are well known~\cite{Strelets2001,Menter2011}.
The low-order method presented in this paper uses input $C_l$, $C_d$, $C_m$, and $f$ vs. $\alpha$ curves obtained from 2D RANS CFD to make predictions for 3D wings.
While the low-order predictions compare well with 3D RANS CFD solutions, it is prudent to verify the results using 2D input data  and 3D solutions obtained from a higher fidelity computational method such as LES or DES in a future study.
% While the low-order results obtained in this work using 2D RANS solutions as input match excellently with 3D RANS solutions, it would be prudent to verify the results using a higher fidelity computational method such as LES or DES in future studies. For the scope of this study, the

\subsection{Low aspect ratio wings}
\revnote{
As noted in \sref{sec:unswept-results}, the predictions from the low-order
method become less accurate as the aspect ratio of the wing reduces.
At high angles of attack, the tip vortex separates from
the wingtip which affects the forces and moments on the
wing.
While the detached vortex does not significantly affect
the flow over wings of higher aspect ratios,
low aspect ratio wings ($\ar \lessapprox 6$) experience a
greater disruption of
flow and therefore modeling the effects of the modified
wingtip vorticity is essential to accurate modeling of
the flow and loads on such wings.
The current method can be easily augmented with a wingtip
vortex model, such as that proposed by Loewenthal and
Gopalarathnam \cite{loewenthal2019}.}{\#3.3}

\subsection{Airfoils exhibiting sharp stall}
% \PHNote{We could demonstrate what happens with the Merz wing and the NACA0009 wing?}
The current implementation of the \methodabbr also has problems converging at post-stall conditions for wings with airfoils that have abrupt stall behavior (sudden drop in $C_l$ at stall). This convergence problem is a subject of continuing work, and may require improvements to the numerical methods used in this work.

 \section{Conclusions}
\label{sec:ch2-summary}

A traditional vortex lattice method (VLM) provides accurate potential-flow solutions for three-dimensional wings at low angles of attack at which the boundary layer is thin and mostly attached. At higher angles of attack, at which there is significant flow separation, a VLM significantly overpredicts the lift and moment produced by the wing. The nonlinear decambering method presented in this paper is an augmentation to the potential-flow VLM to model the effects of boundary-layer separation at near-stall and post-stall angles of attack. Viscous data for the two-dimensional airfoils, in the form of experimentally- or computationally-obtained lift, drag, and moment curves are provided as inputs to the decambering method. For each wing section, the deviations in the lift and moment coefficients are computed as the differences in values obtained from the viscous input curves at the effective angle of attack for the section and those predicted by the VLM for that section. A parabolic decambering flap is deflected from the separation location at each section with the aim of bringing these deviations to zero for all the wing sections. The iterative procedure to determine the decambering-flap shapes for all the sections converges when the operating point for each section lies on its two-dimensional viscous input curve. 

The total loads predicted by the low-order method agree well with experimental results and computational solutions for a variety of unswept wings. Stall angle, and lift and moment coefficients at stall are slightly overpredicted in some cases, but generally follow the trends seen in the experimental and computational results. The predictions from the method are seen to become less accurate for wings with low aspect ratio, which is attributed to the unmodeled effects of the separated and rolled-up wing tip vortices. Spanwise distributions of lift and moment compare well with CFD solutions even at post-stall conditions. The method correctly predicts the stall characteristics of unswept wings, with root sections stalling before tip sections for rectangular wings, and stall occurring at the outboard sections first on tapered planforms. A unique capability of the current method is to predict the spanwise variation of the flow separation on the wing. While the method is unable to resolve stall cells that occur on unswept wings at high angles of attack, the predicted separation patterns generally agree well with those obtained from skin-friction lines calculated using CFD solutions. The shapes of the decambering flap also closely mimic the shapes of the separated boundary layer at various sections of a stalled wing. For the wings experiencing a small roll rate, the method accurately predicts the roll damping at pre-stall angles of attack, and the loss thereof after stall.

Improvements to the method could focus on improved convergence for airfoils having abrupt stall characteristics, extensions to very high post-stall angles of attack, and the capability to handle swept wings. Nevertheless, even in its current state, the method shows promise for rapid prediction of stall behavior of unswept wings in steady flight and with quasi-steady roll rates, providing useful capability for design, modeling, and simulation at post-stall conditions.

\section*{Acknowledgements}
This research effort was supported by Master Subaward Agreement \#C15-2B00NCSU (Base) under NASA Cooperative Agreement to NIA \#NNL09AA00AA 08-01-2015 from the NASA Langley Research Center under the Vehicle Systems Safety Technologies project. We thank technical monitors Gautam Shah and Neal Frink of NASA Langley for their support and collaboration.

\bibliography{refs}

\end{document}